\documentclass[structabstract]{aa}
\usepackage{graphicx}
\usepackage{txfonts}
\usepackage{natbib}
\usepackage{color}
\usepackage{longtable} %
\usepackage{supertabular}
\usepackage{hyphenat}
\usepackage{upgreek}
\usepackage{fmtcount}
\usepackage{ulem}
\usepackage{subfig}
\usepackage{longtable}
\usepackage{xspace}

\bibpunct{(}{)}{;}{a}{}{,} % to follow A&A style

\newcommand{\twCO}{$^{12}$CO\xspace}
\newcommand{\thCO}{$^{13}$CO\xspace}

\newcommand{\CeiO}{C$^{18}$O\xspace}
\newcommand{\CseO}{C$^{17}$O\xspace}
\newcommand{\HOCp}{HOC$^+$\xspace}

\newcommand{\HCOp}{HCO$^+$\xspace}

\newcommand{\mum}{$\upmu$m\xspace}

\newcommand{\XCO}{$X_{\rm CO}$\xspace}
\newcommand{\COU}{\,$10^{20}$ (K km s$^{-1}$)$^{-1}$ cm$^{-2}$\xspace}
\newcommand{\lsun}{$L_{\odot}$\xspace}

\newcommand{\LFIR}{$L_{\rm FIR}$\xspace}
\newcommand{\msun}{M$_{\odot}$\xspace}

\newcommand{\cmt}{cm$^{-3}$\xspace}

\newcommand{\Gn}{$G_{\rm 0}$\xspace}
\newcommand{\Av}{$A_{\rm v}$\xspace}
\newcommand{\nh}{$n_{\rm H}$\xspace}
\newcommand{\Ha}{H$\upalpha$\xspace}

\newcommand{\kms}[0]{\,km\,s$^{-1}$\xspace}

\newcommand{\CII}{[\ion{C}{ii}]\xspace}

\newcommand{\HII}{\ion{H}{ii}\xspace}
\newcommand{\HI}{\ion{H}{i}\xspace}

\newcommand{\herschel}{{\it{Herschel}}\xspace}

\newcommand{\hermes}{{\tt HerM33es}\xspace}

\newcommand{\maxHCNCO}{2.9}
\newcommand{\minHCNCO}{0.4}
\newcommand{\meanHCNCO}{1.5}
\newcommand{\stdHCNCO}{0.8}

\newcommand{\maxHCOpCO}{3.5}
\newcommand{\minHCOpCO}{0.6}
\newcommand{\meanHCOpCO}{1.9}
\newcommand{\stdHCOpCO}{1.0}

\begin{document}

\title{Dense gas in M\,33 (\hermes)\thanks{Based on observations with the IRAM
    30m telescope, Herschel, and other observatories. IRAM is supported by
    CNRS/INSU (France), the MPG (Germany), and the IGN (Spain). \herschel is an
    ESA space observatory with science instruments provided by European-led
    Principal Investigator consortia and with important participation from
    NASA.}$^{,}$\thanks{FITS files of the presented spectra of the ground-state
    transitions of HCN, \HCOp, \twCO and \thCO are available at the CDS via
    anonymous ftp to cdsarc.u-strasbg.fr (130.79.128.5) or via
    http://cdsweb.u-strasbg.fr/cgi-bin/qcat?J/A+A/.}}  \subtitle{}

\author{ C.\,Buchbender\inst{1} \and C.\,Kramer\inst{1} \and
  M.\,Gonzalez-Garcia\inst{1} \and F.\,P.\,Israel\inst{2}\and
  S.\,Garc\'{i}a-Burillo\inst{3}\and P.\,van der Werf\inst{4}\and
  J.\,Braine\inst{5}\and E.\,Rosolowsky\inst{6}\and B.\,Mookerjea\inst{7}\and
  S.\,Aalto\inst{8}\and M.\,Boquien\inst{9}\and P.\,Gratier\inst{10}\and
  C.\,Henkel\inst{11}\inst{15}\and G.\,Quintana-Lacaci\inst{12} \and
  S.\,Verley\inst{13}\and F.\,van der Tak\inst{14} }

\institute{Instituto Radioastronom\'{i}a Milim\'{e}trica, Av. Divina Pastora
  7, Nucleo Central, E-18012 Granada, Spain \\\email{buchbend@iram.es} %1
  \and Sterrewacht Leiden, Leiden University, PO Box 9513, 2300 RA Leiden, The
  Netherlands %2
  \and Observatorio Astron\'{o}mico Nacional (OAN)-Observatorio de Madrid,
  Alfonso XII, 3, 28014-Madrid, Spain %3
  \and Leiden Observatory, Leiden University, P.O. Box 9513, NL-2300 RA Leiden,
  The Netherlands %4
  \and Laboratoire d'Astrophysique de Bordeaux, Universit\'{e} de Bordeaux,
  OASU, CNRS/INSU, 33271 Floirac, France %5
  \and University of British Columbia Okanagan, 3333 University Way, Kelowna BC
  V1V 1V7 Canada %6
  \and Department of Astronomy \& Astrophysics, Tata Institute of Fundamental
  Research, Homi Bhabha Road, Mumbai 400005, India %7
  \and Department of Earth and Space Sciences, Chalmers University of
  Technology, Onsala Observatory, 439 94 Onsala, Sweden %8
  \and Laboratoire d'Astrophysique de Marseille - LAM, Universi\'{e}
  Aix-Marseille \& CNRS, UMR7326, 38 rue F. Joliot-Curie, 13388 Marseille CEDEX
  13, France %9
  \and IRAM, 300 rue de la Piscine, 38406 St. Martin d'H\`{e}res, France %10
  \and Max-Planck Institut f\"{u}r Radioastronomie (MPIfR), Auf dem H\"{u}gel
  69, 53121 Bonn, Germany %11
  \and Departamento de Astrof\'{i}sica, Centro de Astrobiolog\'{i}a, CSIC-INTA,
  Ctra.\ de Torrej\'{o}n a Ajalvir km 4, 28850 Madrid, Spain %12
  \and Dept.\ F\'{i}sica Te\'{o}rica y del Cosmos, Universidad de Granada,
  Spain
  % 13
  \and SRON Netherlands Institute for Space Research, Landleven 12, 9747 AD
  Groningen, The Netherlands %14
  \and Astron. Dept., King Abdulaziz University, P.O. Box 80203, Jeddah, Saudi
  Arabia } \offprints{C.\,Buchbender, \email{buchbend@iram.es}} \date{}

\abstract
{}
% leave it empty if necessary aims heading (mandatory)
{ We aim to better understand the emission of molecular tracers of the diffuse
  and dense gas in giant molecular clouds and the influence that
  metallicity, optical extinction, density, far-UV field, and star formation
  rate have on these tracers.}
{ Using the IRAM 30\,m telescope, we detected HCN, \HCOp, \twCO, and \thCO in
  six GMCs along the major axis of M\,33 at a resolution of $\sim$\,114\,pc and
  out to a radial distance of 3.4\,kpc. Optical, far-infrared, and
  submillimeter data from Herschel and other observatories complement these
  observations. To interpret the observed molecular line emission, we created
  two grids of models of photon-dominated regions, one for solar and one for
  M\,33-type subsolar metallicity.}
{ The observed \HCOp/HCN line ratios range between 1.1 and 2.5.  Similarly high
  ratios have been observed in the Large Magellanic Cloud. The HCN/CO ratio
  varies between 0.4\% and 2.9\% in the disk of M\,33. The \twCO/\thCO\ line
  ratio varies between 9 and 15 similar to variations found in the diffuse gas
  and the centers of GMCs of the Milky Way. Stacking of all spectra allowed HNC
  and C$_2$H to be detected. The resulting \HCOp/HNC and HCN/HNC ratios of
  $\sim$ 8 and 6, respectively, lie at the high end of ratios observed in a
  large set of (ultra-)luminous infrared galaxies. HCN abundances are lower in
  the subsolar metallicity PDR models, while \HCOp abundances are enhanced. For
  HCN this effect is more pronounced at low optical extinctions. The observed
  \HCOp/HCN and HCN/CO line ratios are naturally explained by subsolar PDR
  models of low optical extinctions between 4 and 10\,mag and of moderate
  densities of $n$\,$\,3\,10^3$--$\,3\,10^4$\,cm$^{-3}$, while the FUV field
  strength only has a small effect on the modeled line ratios. The line ratios
  are almost equally well reproduced by the solar-metallicity models,
  indicating that variations in metallicity only play a minor role in
  influencing these line ratios.}
{}



\maketitle

% ________________________________________________________________

   \begin{figure}[ht] %CKr
     \centering
     \includegraphics[width=0.48\textwidth, angle=0]{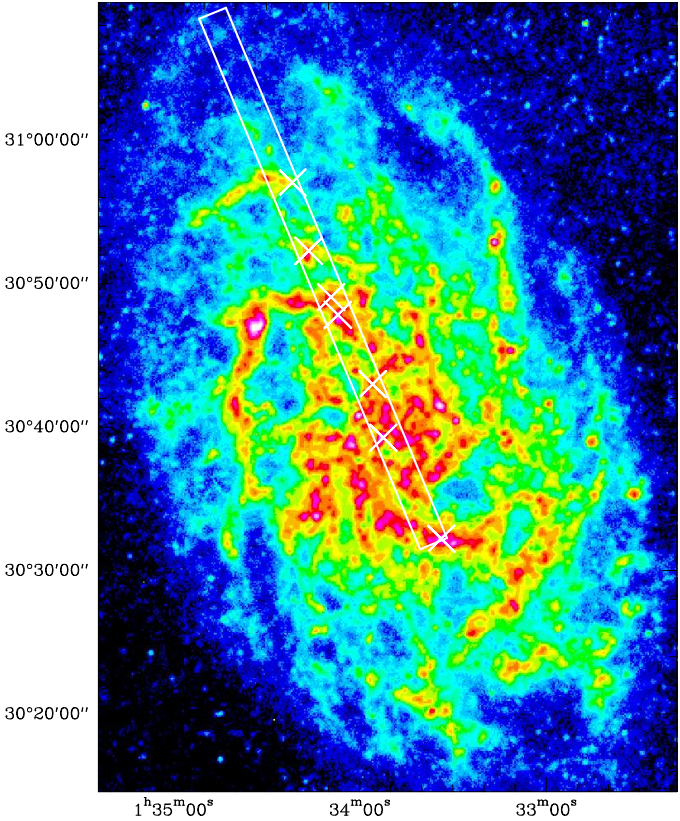}
     \caption{SPIRE 250$\,\upmu$m map of M\,33 \protect\citep{Xilouris2012}.
       The rectangle delineates the $2'\times40'$ wide strip along the major
       axis shown in Fig.\,2. Crosses mark the positions of the observed GMCs.}
     \label{fig-spire250}
   \end{figure}

   \section{Introduction}

   Owing to their large dipole moments, even the rotational ground state
   transitions of HCN and \HCOp trace dense molecular gas with densities in
   excess of $\sim$\,$10^4$\,cm$^{-3}$. Because stars condense out of dense cores
   of giant molecular clouds (GMCs), both molecules are promising tracers of
   star formation (SF) and the star formation rate (SFR). A series of papers
   \citep{Gao2004a, Gao2004b, Wu2005, Gao2007, Baan2008, Gracia-Carpio2008,
     Wu2010, Garcia-Burillo2012, Liu2012} have recently investigated the
   correlation of HCN (and partly \HCOp) with far-infrared (FIR) luminosities
   (\LFIR) in galactic GMCs, centers of nearby galaxies, and
   \mbox{(ultra-)luminous} galaxies (LIRGs/ULIRGs), showing that HCN is indeed
   a good tracer of SF and tightly correlated with \LFIR. There are, however,
   findings that complicate this picture \citep[see e.g.][]{Costagliola2011}.  In
   contrast to CO, which traces the bulk of the molecular gas-phase
   carbon, HCN and \HCOp are minor species. Their abundances are therefore more
   strongly influenced by the details of the chemical network
   \citep[e.g.][]{Lopez-Sepulcre2010}. \HCOp is linked via ion-molecule
   reactions to the ionization equilibrium. Its collisional cross-section is
   close to a factor $10$ larger than that of HCN, which is linked to the
   hydrocarbon chemistry and the amount of nitrogen in the gas phase.
   Elemental depletion in low-metallicity environments may therefore have a
   strong effect on its abundance.

   Most extragalactic observations of HCN and \HCOp have so far been restricted
   to the nuclei of galaxies or their integrated fluxes. Exceptions are
     a study of HCN and \HCOp emission in the disk of M\,31 by
     \citet{Brouillet2005}, LMC observations
     \citep{Chin1996,Chin1997,Chin1998,Heikkilae1999}, HCN maps of seven
     Seyfert galaxies by \citet{Curran2001}, and also HCN mapping along the
     major axis in 12 nearby galaxies by \citet{Gao2004a, Gao2004b} (hereafter
     GS04a,b). We also recommend the studies of HCN and CO and their ratios in M51 by
     \citet{Kuno1995} and \citet{Schinnerer2010}. The interstellar medium
   (ISM) of nuclear regions are often subject to particularly strong heating
   sources because they are often dominated by starbursts with intense UV fields
   heating the gas or active galactic nuclei (AGNs) with strong X-ray
   emission. Indeed, the HCN to \HCOp line intensity ratios are found to be
   systematically higher in AGN-dominated regions, such as in the central part
   of NGC\,1068, and low in pure starburst environments, as in M82
   \citep[e.g.][]{Kohno2003, Imanishi2009, Krips2008, Krips2011}.

   We have targeted seven GMCs along the major axis of M\,33 out to a radial
   distance of 4.6\,kpc, using the IRAM 30\,m telescope. M\,33 is a spiral
   galaxy with Hubble type SA(s)cd located at a distance of only 840\,kpc
   (Table~\ref{Tab:M33Properties} and Fig.~\ref{fig-spire250}). It is the third
   largest member of the Local Group (after M\,31 and the Milky
   Way). Observations of small-scale structures in M\,33 do not suffer from
   distance ambiguities as galactic observations do. Its small distance allows
   us to obtain a spatial resolution of $\sim$\,114\,pc (i.e. 28\,\arcsec) at a
   frequency of 89\,GHz (i.e. 3.3\,mm) with the 30\,m telescope. M\,33 is seen
   at an intermediate inclination of $i=56^{\circ}$, yielding a short
   line-of-sight depth, which allows us to study individual cloud complexes. It
   is roughly ten times less massive than the Milky Way, and its overall
   metallicity is \mbox{12+log O/H=8.27}, subsolar by about a factor two
   \citep{Magrini2010}. Therefore M\,33 is particularly interesting to compare
   with the Milky Way, but also with the Large Magellanic Cloud that has a
   metallicity similar to M\,33 \citep{Hunter2007}.

   Using the IRAM 30\,m telescope, \citet{Rosolowsky2011} (hereafter RPG11)
   observed four massive GMCs of more than $3\,10^5$\,\msun\ in M\,33,
   searching for the ground-state transition HCN\@. They detected HCN in only
   two of the GMCs. The observed GMCs are under-luminous in HCN by factors
   between two to seven relative to their CO emission when compared to averaged
   values in the Milky Way.

\begin{figure*}[ht]
  \centering
  \includegraphics[width=0.99\textwidth]{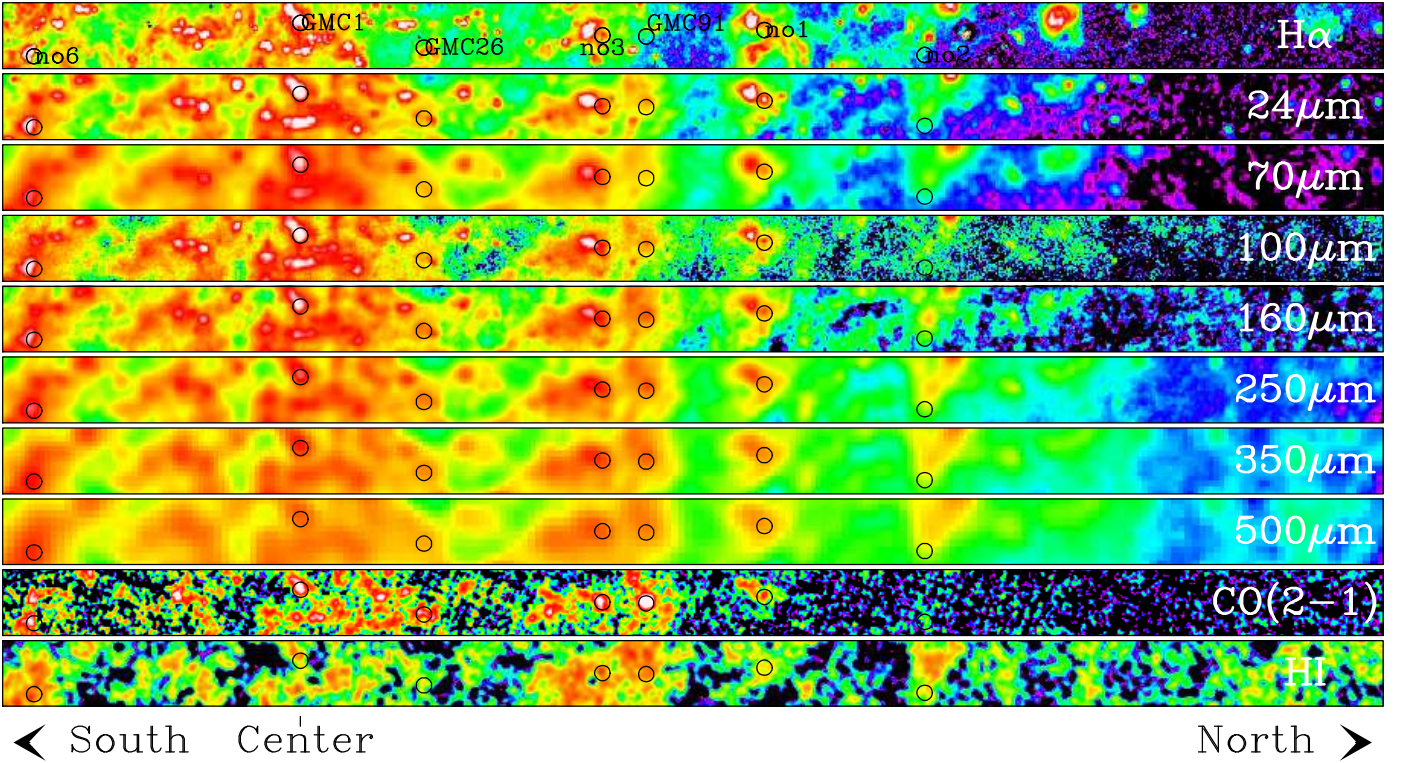}
  \caption{Observed positions towards seven giant molecular clouds (GMCs)
    within a $2'\times40'$ strip along the major axis of M\,33.  The strip
    extends from $10'$ south of the galactic center to $33.3'$ north. The
    center of the strip is at 01:34:11.8 +30:50:23.4 (J2000). Circles indicate
    the 30\,m beam size of $28''$ at 90\,GHz. Panels show from top to
      bottom: integrated intensities of H$\upalpha$ emission \protect\citep{Hoopes2000}
      24; 70\,$\upmu$m emission observed with Spitzer \protect\citep{Tabatabaei2007};
      continuum emission between 100$\,\upmu$m and 500$\,\upmu$m observed with
      PACS and SPIRE in the framework of the {\tt HerM33es} program
      \protect\citep{Kramer2010, Boquien2011, Xilouris2012}; \mbox{\twCO\,2--1} 30\,m
      observation and \HI VLA data, both taken from \protect\citet{Gratier2010}. All
    data are shown at their original resolutions.}
  \label{Fig:allp}%
\end{figure*}

   Here, we present new, deep observations of the ground-state transitions of
   HCN, \HCOp, \thCO, and CO towards seven GMCs in M\,33 including three of the
   clouds observed by RPG11. All four tracers are detected in six of the
   GMCs. The relative weakness of HCN emission is confirmed and interpreted
   using models of photon-dominated regions (PDRs). To better characterize the
   observed GMCs, we estimated their star formation rate, total infrared
   luminosities, and FUV fields using a large ancillary data set compiled in
   the framework of the Herschel open time key project {\tt HerM33es}
   \citep{Kramer2010}. Figure~\ref{Fig:allp} shows a subset of this data set.

   % +_+_+_+_+_+_+_+_+_+_+_+_+_+_+_+_+_+_+_+_+_+_+_+_+_+_+_+_+_+_+_+_+

   \begin{table}[h*]
     \caption[]{\label{Tab:M33Properties} Basic properties of M\,33}
     \begin{center}
       \begin{tabular}{lrl}
	 \hline \hline
	 & M\,33 & References \\
	 \noalign{\smallskip} \hline \noalign{\smallskip}
	 RA(2000)                     & 01:33:51.02 \\
	 DEC(2000)                    & 30:39:36.7 \\
	 Type                         & SA(s)cd & 1 \\
	 Distance [kpc]               & 840 & 2 \\
	 $11''$ (30m @ 230\,GHz) equal to     & 45\,pc \\
	 $21''$ (30m @ 115\,GHz) equal to     & 86\,pc \\
	 $28''$ (30m @ 89\,GHz) equal to      & 114\,pc \\
	 LSR velocity [km\,s$^{-1}$]  & $-180$ \\
	 Position Angle [deg]         & 22.5 & 3\\
	 Inclination [deg]            & 56 & 4\\
	 $R_{25}$                        & $30.8'$ or 7.5\,kpc \\
	 % $L_{\rm tot}$ & \\
	 % $M_{\rm tot}$ & \\
	 \noalign{\smallskip} \hline \noalign{\smallskip}
       \end{tabular}
       \tablebib{(1)~\citet{Devaucouleurs1991};
	 (2)~\citet{Galleti2004,Freedman1991}; (3)~\citet{Paturel2003};
	 (4)~\citet{Regan-vogel1994,Zaritsky1989}.}
     \end{center}
   \end{table}

%%%%%%%%%%%%%%%%%%%%%%%%%%%%%%%%%%%%%%%%%%%%%%%%%%%%%%%%%%%%%%%%%%%%%%%%%%%%%%

\begin{figure*}[ht!]
  \centering
  \includegraphics[width=0.8\textwidth]{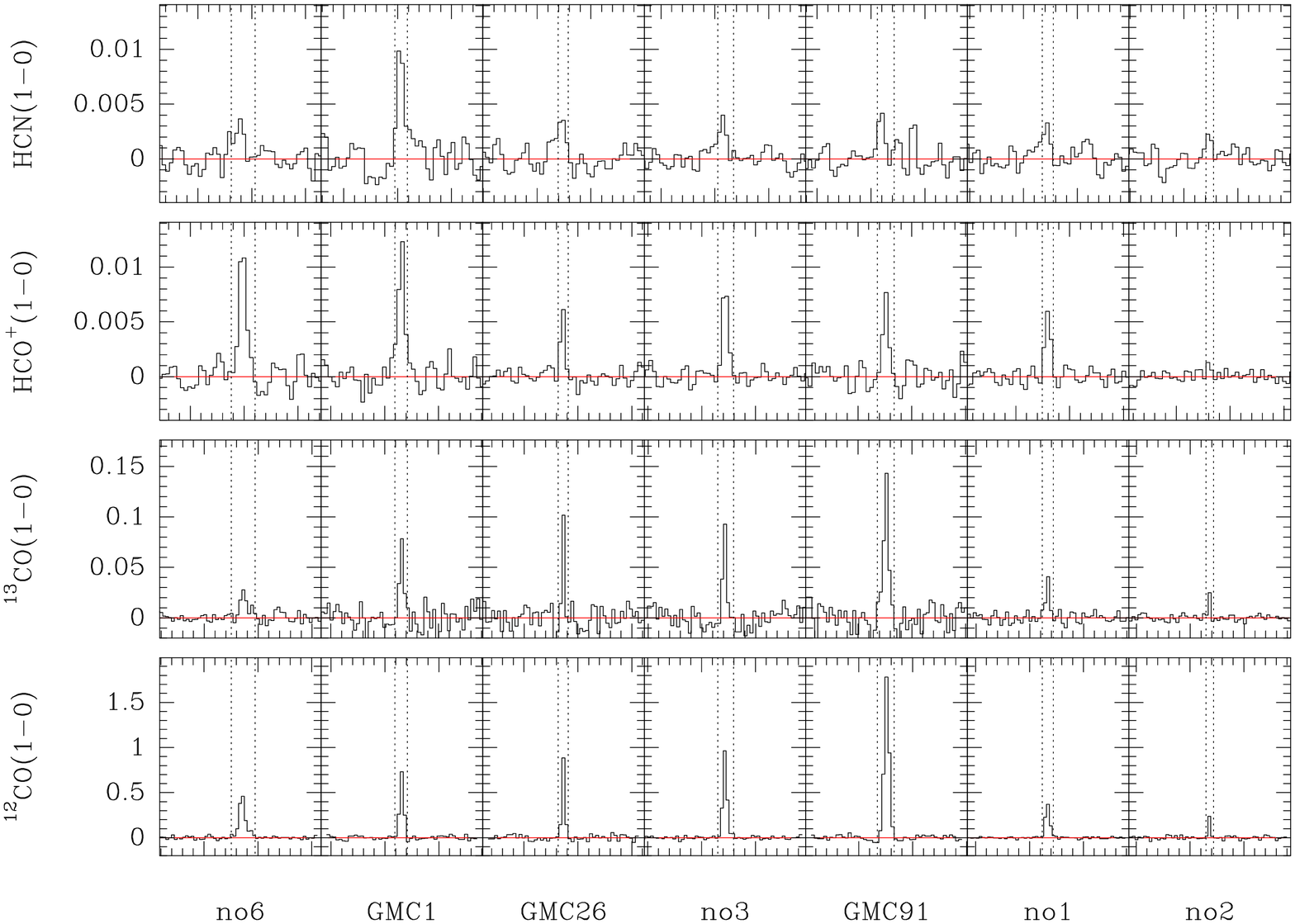}
  \caption{Spectra of the ground-state transitions of HCN, \HCOp, \twCO, and
    \thCO at the positions of seven GMCs along the major axis of M\,33
    (cf. \mbox{Fig.\,1, 2)}. \twCO and \thCO have been observed in
    position switching; HCN and \HCOp with wobbler switching. The spectra are
    shown on a main beam brightness temperature scale ($T_{\rm mb}$). The
    velocity resolution is given by the spectrometer with the lowest
    resolution, i.e. WILMA, and is 5.4\,km\,s$^{-1}$ in case of \twCO\ and
    \thCO, and 6.7\,km\,s$^{-1}$ for HCN and \HCOp. Center velocities are
    listed in Table~\ref{table:values}. The local standard of rest ({\tt lsr})
    velocity displayed covers 300\,km\,s$^{-1}$. The same velocity range is
    used to determine the baseline (red lines), excluding the line windows
    determined from \mbox{\twCO\,1--0} (dotted lines),
    cf. Fig.~\ref{Fig:COFull}}
  \label{fig-lines}
\end{figure*}

\section{IRAM 30\,m observations}
\label{observations}

We used the IRAM 30\,m telescope to perform single-pointed observations of the
ground-state transitions of HCN, \HCOp, \twCO, and \thCO towards seven GMCs in
M\,33. Observations were carried out between December 2008 and July 2012,
comprising a total of 109\,hours of observing time. In 2008, we used the now
decommissioned A100 and B100 receivers that had a bandwidth of 500\,MHz and the
1MHz filterbank, to observe each of the four lines individually.

The bulk of the observations were carried out in 2009 employing the new
eight-mixer receiver EMIR and its instantaneous bandwidth of 16\,GHz in each
polarization, connected to the wide-band WILMA autocorrelator backend with
2\,MHz spectral resolution. This setup allowed simultaneous observation of HCN
with \HCOp and \twCO with \thCO. One advantage of the simultaneous observations
is that the relative intensity calibration of the lines is very accurate. The
observations were carried out in wobbler switching mode using the maximum
available throw of $\pm$\,120\,\arcsec and a switching frequency of two
seconds. This mode ensures more stable baselines than the position-switching
mode. However, the velocity resolution of about 6\,kms$^{-1}$ in the 3\,mm band
only barely resolves the spectral lines of M\,33, which are typically
10--15\,km\,s$^{-1}$ wide \citep{Gratier2010}. The beam sizes are $21''$ at
115\,GHz and $28''$ at 89.5\,GHz, corresponding to a spatial resolution of
86\,pc and 114\,pc, respectively, in M\,33 (cf.
Table~\ref{Tab:M33Properties}).

The observations of \twCO and \thCO were repeated in June and July 2012 using
position-switching and the FTS spectrometers with an off position outside of
the disk of M\,33 to exclude the possibility of self-chopping effects in the
spectra. The latter were present in some of the earlier wobbler-switched \twCO
spectra. Due to the high critical density of the dense gas tracers HCN and
\HCOp, as well as the observed velocity gradient along the major axis of M\,33
(see Table~\ref{table:values}), we reckon that self-chopping is less probable
for these lines. Please note that the \twCO data of GMC1, GMC91, and GMC26 are
taken from RPG11 who also used position-switching.

All data were reduced using the GILDAS
\footnote{http://www.iram.fr/IRAMFR/GILDAS} software package. Each scan was
inspected and scans with poor baselines or unreasonably high rms values were
rejected. Before averaging, linear baselines were fitted and removed. The data
were regridded to a common velocity resolution. Spectra were converted from
the $T_{\rm A}^*$ to the $T_{\rm mb}$ scale by multiplying with the ratio of
forward efficiency ($F_{\rm eff}$~=~95\%) to main beam efficiency ($B_{\rm
  eff}$~=~81\%), taken to be constant for the observed 3\,mm lines. The reduced
spectra are shown in \mbox{Fig.~\ref{fig-lines}}.

Integrated intensities were extracted from the spectra on a $T_{\rm mb}$ scale
by summing all channels inside a velocity range around each particular
line. The velocity range was determined by eye for each position from the full
width to zero intensity (FWZI) of the \mbox{\twCO\,1--0} line and is marked in
\mbox{Fig.~\ref{fig-lines}}. We determined $\sigma$ uncertainties of the
integrated intensities by measuring the baseline rms ($T^{\rm rms}_{\rm mb}$)
in a 300\,km\,s$^{-1}$ window centered on the particular line using the
corresponding \mbox{\twCO\,1--0} FWZI as baseline window and using $\sigma =
T^{\rm rms}_{\rm mb}\,\sqrt{N}\,\Delta v_{\rm res}$ with the number of channels
$N$ and the channel width $\Delta v_{\rm res}$. In case the 1\,$\sigma$ value is
higher than the typical 10\,\% calibration error of the IRAM\,30\,m, the former
is used to estimate the observational error for the following analysis. If the
integrated intensities are lower than~3\,$\sigma$, we use this value as an
upper limit. Table\,\ref{table:values} lists the observed intensities,
intensity ratios, and further ancillary data. For details on the latter see
Appendix~\ref{AppendixA}.  Error estimates are given in parentheses after the
integrated intensities in Table~\ref{table:values}.

%%%%%%%%%%%%%%%%%%%%%%%%%%%%%%%%%%%%%%%%%%%%%%%%%%%%%%%%%%%%%%%%%%%%%%%%%%%

\section{GMCs: Selection of positions and properties}

Motivated by the \hermes project, the GMCs were selected to lie within a
$2'\times40'$ wide strip along the major axis of M\,33 shown in
Figs.~\ref{fig-spire250}~and~\ref{Fig:allp} at a range of galacto-centric
distances of up to 4.6\,kpc. Three of the GMCs (GMC1, GMC26, GMC91) belong to
the sample of CO-bright clouds studied by RPG11 in search of HCN emission. We
added four other GMCs (no6, no3, no1, no2) to increase the range of studied
galacto-centric radii, as well as physical conditions.

Table~\ref{table:values} lists their observed properties and
Appendix~\ref{AppendixA} describes in detail how they were derived. The masses
of the molecular gas traced by CO, calculated using $X_{\rm CO}$-factors
derived individually for every cloud as a function of integrated
\mbox{CO\,1--0} intensities and total IR luminosity
(cf. Appendix~\ref{XCODer}), vary by a factor 130 between $0.1\,10^5$ (GMC no2)
and $13\,10^5$\,\msun\ (GMC1). The SFRs vary by more
than a factor 50 and the far ultraviolet (FUV) field strengths by a factor
larger than 20.  The GMC near the nucleus, GMC1, is the most massive and shows
the strongest SFR, as well as the highest FUV flux, while GMC\,no2 at 4.6\,kpc
radial distance is the least massive in molecular mass and shows only weak
activity.

Individual areas of the strip have been mapped in \CII\ and other FIR lines in
the framework of the {\tt HerM33es} project, which will yield additional
insight into the properties of the ISM of M\,33. In the first papers on \CII,
we focused on the \HII\ region BCLMP\,302 \citep{Mookerjea2011} which is
associated with GMC\,no3, and on BCLMP\,691 (Braine et al., in subm.), which
lies near GMC\,no1.

\citet{Gratier2012} identified over three hundred \mbox{\twCO\,2--1} clumps in
M\,33 and present a detailed view of each individual clump in \Ha, 8\,\mum,
24\,\mum, and FUV, together with the corresponding HI and \mbox{\twCO\,2--1}
spectra and further complementary data. The seven GMCs discussed here are among
the identified clumps. In Table~\ref{table:values} we give the corresponding
clump numbers.

%%%%%%%%%%%%%%%%%%%%%%%%%%%%%%%%%%%%%%%%%%%%%%%%%%%%%%%%%%%%%%%%%%%%%

\section{Observed line ratios}

\begin{table*}
  \caption{Observed intensities and complementary data}
  \label{table:values}
  \centering

\begin{tabular}{cccccccc}
\hline
\hline
    &	no6	&	GMC\,1	&	GMC\,26	&	no3	&	GMC\,91	&	no1	&	no2\\
\hline
Clump number \tablefootmark{a}	&	42	&	108	&	128	&	256	&	245	&	300	&	320\\
RA [J2000]	&	01:33:33.77	&	01:33:52.40	&	01:33:55.80	&	01:34:07.00	&	01:34:09.20	&	01:34:16.40	&	01:34:21.77\\
DEC [J2000]	&	+30:32:15.64	&	+30:39:18.00	&	+30:43:02.00	&	+30:47:52.00	&	+30:49:06.00	&	+30:52:19.52	&	+30:57:4.99\\
$V_{\rm LSR}$ [km\,s$^{-1}$]	&	-133.0	&	-168.0	&	-227.0	&	-257.0	&	-247.0	&	-266.0	&	-264.0\\
$R$ [kpc]	&	2.01	&	0.11	&	0.873	&	2.18	&	2.51	&	3.38	&	4.56\\
$I_{\rm ^{12}CO(1-0)}$ [K\,km\,s$^{-1}$]	&	7.1 (10\,\%)    &	7.2
(10\,\%) \tablefootmark{b}	&	6.9 (10\,\%) \tablefootmark{b}	&	9.4
(10\,\%)	&	21.6 (10\,\%) \tablefootmark{b}	&	4.0 (10\,\%)	&	1.4 (10\,\%)\\
FWHM $^{12}$CO(1-0) [\kms] \tablefootmark{c}	&	11 (0.5)	&	8 (0.2)	&	6 (0.2)	&	8 (0.1)	&	11 (0.1)	&	9 (0.2)	&	4 (0.2)\\
$I_{\rm ^{12}CO(2-1)}$ [K\,km\,s$^{-1}$] \tablefootmark{d}	&	8.9 (15\,\%)	&	10.6 (15\,\%)	&	7.0 (16\,\%)	&	9.4 (15\,\%)	&	19.3 (15\,\%)	&	6.2 (15\,\%)	&	0.7 (15\,\%)\\
$I_{\rm ^{13}CO(1-0)}$ [mK\,km\,s$^{-1}$]	&	468 (12\,\%)	&	799 (13\,\%)	&	541 (19\,\%)	&	772 (13\,\%)	&	1690 (10\,\%)	&	369 (12\,\%)	&	132 (15\,\%)\\
$I_{\rm HCO^+(1-0)}$ [mK\,km\,s$^{-1}$]	&	205 (10\,\%)	&	182 (10\,\%)	&	66 (10\,\%)	&	119 (10\,\%)	&	97 (15\,\%)	&	77 (10\,\%)	&	$<$ 12\\
$I_{\rm HCN(1-0)}$ [mK\,km\,s$^{-1}$]	&	82 (20\,\%)	&	164 (10\,\%)	&	56 (16\,\%)	&	61 (16\,\%)	&	67 (24\,\%)	&	56 (17\,\%)	&	26 (28\,\%)\\
$I_{\rm HNC(1-0)}$ [mK\,km\,s$^{-1}$]	&	$<$ 43.9	&	$<$ 44.8	&	$<$ 15.6	&	$<$ 22.3	&	$<$ 47.5	&	$<$ 25.9	&	$<$ 20.4\\
rms [mK] \tablefootmark{e}	&	1.0	&	1.4	&	1.0	&	0.7	&	1.1	&	0.9	&	0.8\\
\hline & & & & & & \\
%BFF 12\arcsec $\rightarrow$ 22\arcsec	&	1.8	&	1.6	&	1.4	&	1.6	&	1.7	&	2.2	&	0.8\\
%BFF 12\arcsec $\rightarrow$ 28\arcsec	&	2.1	&	2.0	&	1.6	&	1.9	&	2.2	&	2.7	&	1.0\\
%BFF 22\arcsec $\rightarrow$ 28\arcsec	&	1.2	&	1.3	&	1.2	&	1.2	&	1.3	&	1.2	&	1.2\\
$I_{\rm HCO^+(1-0)}$/$I_{\rm HCN(1-0)}$	&	2.5 (0.2)	&	1.1 (0.1)	&	1.2 (0.1)	&	1.9 (0.1)	&	1.4 (0.3)	&	1.4 (0.2)	&	$<$ 0.5\\
$I_{\rm HNC(1-0)}$/$I_{\rm HCN(1-0)}$	&	$<$ 0.5	&	$<$ 0.3	&	$<$ 0.3	&	$<$ 0.4	&	$<$ 0.7	&	$<$ 0.5	&	$<$ 0.8\\
$I_{\rm HCN(1-0)}$/$I_{\rm ^{12}CO(1-0)}$ [\%]	&	1.4 (0.3)	&	2.9 (0.4)	&	1.0 (0.2)	&	0.8 (0.2)	&	0.4 (0.1)	&	1.7 (0.4)	&	2.3 (0.7)\\
$I_{\rm HCO^+(1-0)}$/$I_{\rm ^{12}CO(1-0)}$ [\%]	&	3.5 (0.5)	&	3.2 (0.5)	&	1.1 (0.2)	&	1.6 (0.2)	&	0.6 (0.1)	&	2.3 (0.3)	&	$<$ 1.0\\
$I_{\rm ^{12}CO(1-0)}$/$I_{\rm ^{13}CO(1-0)}$	&	15.1 (2.4)	&	9.0 (1.5)	&	12.8 (2.8)	&	12.2 (2.1)	&	12.8 (1.8)	&	10.8 (1.8)	&	10.6 (1.9)\\
%$L_{\rm H\alpha}$ [10$^3$ L$_\odot$]   &	17.1	&	31.1	&	2.1	&	4.4	&	0.8	&	3.3	&	0.4\\
%$\nu_{\rm 24}$L$_{\rm \nu}$(\rm 24) [10$^3$ L$_\odot$]	&	283.7	&	513.6	&	83.5	&	176.5	&	66.7	&	179.2	&	12.6\\
$L^{\prime}_{\rm CO}$ [10$^3$ K\,km\,s$^{-1}$\,pc$^2$]	&	87.7 (8.8)	&	85.4 (8.5)	&	87.9 (8.8)	&	114.6 (11.5)	&	248.9 (24.9)	&	49.8 (5.0)	&	17.3 (1.7)\\
$L_{\rm TIR}$ [10$^6$ L$_{\sun}$]	&	4.4 (0.1)	&	5.9 (0.2)	&	1.4 (0.0)	&	2.7 (0.1)	&	1.3 (0.0)	&	1.7 (0.1)	&	0.3 (0.0)\\
$L_{\rm TIR}$/$L^{\prime}_{\rm HCN}$ [10$^3$]	&	3.5 (0.7)	&	2.4 (0.3)	&	1.6 (0.3)	&	3.0 (0.5)	&	1.3 (0.3)	&	2.0 (0.4)	&	0.8 (0.2)\\
$L_{\rm TIR}$/$L^{\prime}_{\rm HCO^+}$ [10$^3$]	&	1.4 (0.1)	&	2.2 (0.2)	&	1.4 (0.1)	&	1.5 (0.2)	&	0.9 (0.1)	&	1.5 (0.2)	&	$>$ 1.6\\
SFR [M$_{\sun}$ Gyr$^{-1}$ p$c^{-2}$]	&	35.9 (4.3)	&	65.0 (7.8)	&	6.6 (0.8)	&	13.7 (1.8)	&	4.0 (0.6)	&	12.2 (1.7)	&	1.2 (0.1)\\
%$T_{\rm Dust}$ [K]	&	23.4	&	25.7	&	22.1	&	23.0	&	20.9	&	23.1	&	19.7\\
X$_{CO}$	&	5.1	&	6.9	&	1.6	&	3.2	&	1.5	&	2.0	&	0.3\\
$M_{\rm HI}$ [10$^5$ M$_{\sun}$]	&	9.4 (1.4)	&	5.8 (0.9)	&	4.1 (0.6)	&	8.0 (1.2)	&	8.8 (1.3)	&	5.2 (0.8)	&	4.8 (0.7)\\
$M_{\rm H_2}$ [10$^5$ M$_{\sun}$]	&	9.6 (1.0)	&	12.7 (1.3)	&	3.0 (0.3)	&	7.9 (0.8)	&	8.2 (0.8)	&	2.1 (0.2)	&	0.1 (0.0)\\
$G_0$	&	37.3 (1.2)	&	50.7 (1.6)	&	11.6 (0.4)	&	23.5 (0.8)	&	11.3 (0.4)	&	14.4 (0.5)	&	2.5 (0.1)\\
$A_V$	&	6.3 (0.6)	&	6.1 (0.5)	&	2.3 (0.2)	&	5.3 (0.5)	&	5.7 (0.5)	&	2.4 (0.3)	&	1.6 (0.2)\\
\hline
\hline
\end{tabular}
  \tablefoot{Top panel: line intensities are on the $T_{\rm mb}$-scale and on their original
    resolutions: 12\arcsec\ for \mbox{\twCO\,2--1}, 24\arcsec\
    for \twCO\, and \mbox{\thCO\,1--0} and 28\arcsec\ for HCN and
    \mbox{\HCOp\,1--0}. Bottom Panel: line ratios and complementary data are
    on a common resolution of 28\arcsec\xspace. See Appendix~\ref{AppendixA}
    for details.
    \tablefoottext{a}{\twCO\,2--1 clump numbers from \protect\cite{Gratier2012};}
    \tablefoottext{b}{\protect\citet{Rosolowsky2011};}
    \tablefoottext{c}{FWHMs of Gaussian fits to the high resolution \twCO\,1--0
      spectra (Fig.\,\ref{Fig:COFull});}
    \tablefoottext{d}{\protect\citet{Gratier2010};}
    \tablefoottext{e}{Baseline rms of \HCOp spectra at a velocity resolution of 6.7\,\kms.}
  }
\end{table*}

\subsection{Spectra at individual positions: HCO$^+$ and HCN}

\HCOp is detected at six positions with 6 to 12\,mK peak temperatures and with
good signal-to-noise ratios of at least seven; position no2 has not been
detected. HCN emission is detected at the same six positions with
signal-to-noise ratios of four and better; position no2 is but tentatively
detected at a signal-to-noise ratio of 3.5. The \HCOp/HCN ratio of line
integrated intensities of positions where both molecules are detected varies
between 1.1 and 2.5 (Table\,\ref{table:values}). Below, we compare the observed
ratios in detail with ratios found in the Milky Way and in other
galaxies. Although the integrated intensities we find for GMC26, GMC1, and
GMC91 differ up to a factor of two from the values and upper limits given in
RPG11 for the same positions, they are consistent within 3$\sigma$ of the
baseline rms of the observations from RPG11. We attribute the discrepancies to
baseline problems of the RPG11 data.

\subsection{Spectra at individual positions: \twCO and \thCO}

Emission from \twCO and \thCO is detected at all seven positions, though
varying by a factor of more than 15 between GMC\,no2 and GMC91. The ratio of
line integrated \twCO vs. \thCO intensities varies between 9 for GMC\,no1 and
15 for GMC\,no6. In Fig. \ref{Fig:COFull} we show the \twCO and \thCO spectra
at a high resolution of 1\,\kms. The resolved line shapes are Gaussian and the
corresponding FWHMs are given in Table~2.

\subsection{Stacked spectra}

\begin{figure}[ht]

  \subfloat[]{\label{Fig:stackedHCN}
    \includegraphics[width=0.45\textwidth, angle=0]{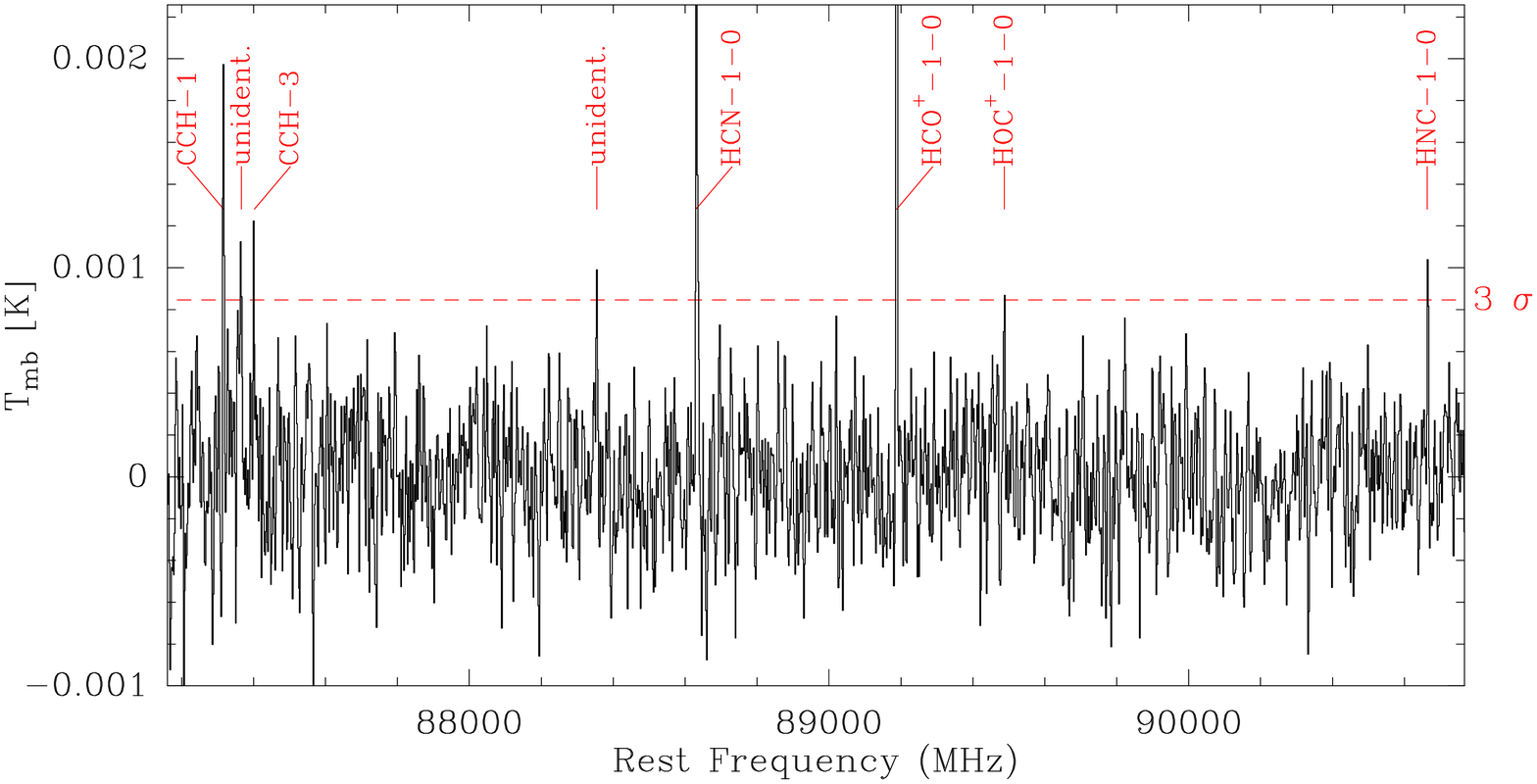}}

  \hskip0.5cm

  \subfloat[]{\label{Fig:stackedCO} \includegraphics[width=0.45\textwidth,
    angle=0]{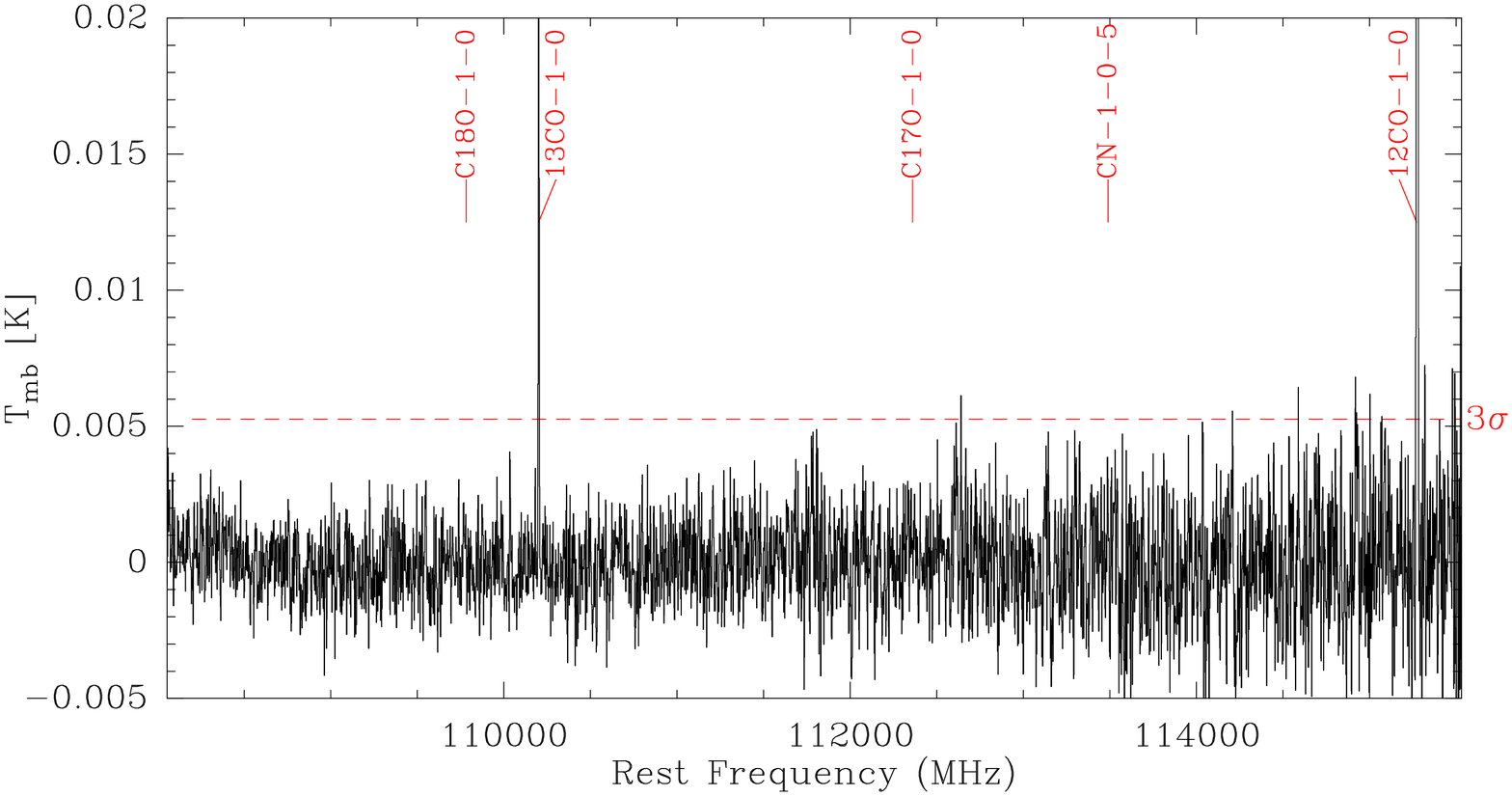}}
  \caption{(a) Stacked spectrum of all data taken in the frequency range
    between 87.2 and 90.8\,GHz. The red dashed line denotes the 3\,$\sigma$
    value average over the entire baseline. The HCN and \HCOp lines are not
    shown up to their maximum peak temperature. (b) Stacked spectrum of the
    wobbler switched data taken in the frequency range between 108.1 and
    115.5\,GHz. The average 3\,$\sigma$ value is shown as red dashed line. The
    baseline noise increases with frequency because of the increasing
    atmospheric opacity.  \CeiO, \CseO, and CN are marked but not detected. The
    \twCO and \thCO lines are not shown up to their maximum peak temperature.}
\end{figure}

\begin{table*}
  \caption{LTE column densities from the stacked spectra.}
  \label{table:stackedTable}
  \centering

\begin{tabular}{cccccccccc}
\hline
\hline
    &	C$_2$H	&	HCN	&	HCO$^+$	&	HOC$^+$	&	HNC	&	$^{13}$CO(1-0)	&	$^{12}$CO(1-0)	&	H$_2$\tablefootmark{a}\\
\hline
$I$ [mK\,km\,s$^{-1}$]	&	25.1	&	81.0	&	110.0	&	$<$ 18.8	&	13.7	&	522.0	&	6280.0	&	-\\
$N$[x]\tablefootmark{b}	&	5.10e+12	&	8.03e+11	&	2.11e+11	&	$<$ 7.10e+10	&	4.18e+10	&	1.80e+15	&	8.55e+15	&	8.47e+20\\
$N$[x]/$N_{\rm H_2}$\tablefootmark{c}	&	-8.40	&	-9.20	&	-9.78	&	$<$ -10.25	&	-10.48	&	-5.85	&	-5.17	&	1.00\\
\hline\end{tabular}

  \tablefoot{
    \tablefoottext{a}{deduced from \thCO, see Appendix~\ref{Appendix:LTE};}
    \tablefoottext{b}{column density;}
    \tablefoottext{c}{logarithmic values.}}
\end{table*}

Stacking of all spectra allows improvement on the signal-to-noise ratios and
detection of faint lines. \mbox{Figure~\ref{Fig:stackedHCN}} shows the stacked
spectrum of all data taken near 89\,GHz. It was created by shifting all spectra
in frequency such that the emission lines align in frequency. Individual
spectra were weighted by integration time. In addition to the lines of HCN and
HCO$^{+}$, the resulting spectrum shows detections of CCH and
\mbox{HNC\,1--0}. The average baseline rms is 0.27\,mK at 5.4\,km\,s$^{-1}$
resolution. \HOCp is tentatively detected with an upper limit of 19
mK\,km\,s$^{-1}$, resulting in a lower limit to the \HCOp/\HOCp ratio of 5.8.

The stacked spectrum centered on 112\,GHz \mbox{(Fig.~\ref{Fig:stackedCO})}
does not show additional detections other than \thCO and \twCO\@ even after
additional smoothing of the velocity resolution. Table~\ref{table:stackedTable}
lists the integrated intensities and upper limits of all detected transitions
in the stacked spectrum, as well as their corresponding LTE column densities and
abundances, derived as explained in Appendix \ref{Appendix:LTE}.

\subsection{Comparison with other sources}

\subsubsection{\HCOp/CO vs. HCN/CO}
\label{co-ratios}
\begin{figure}[ht!]
  \centering
  \includegraphics[width=0.45\textwidth]{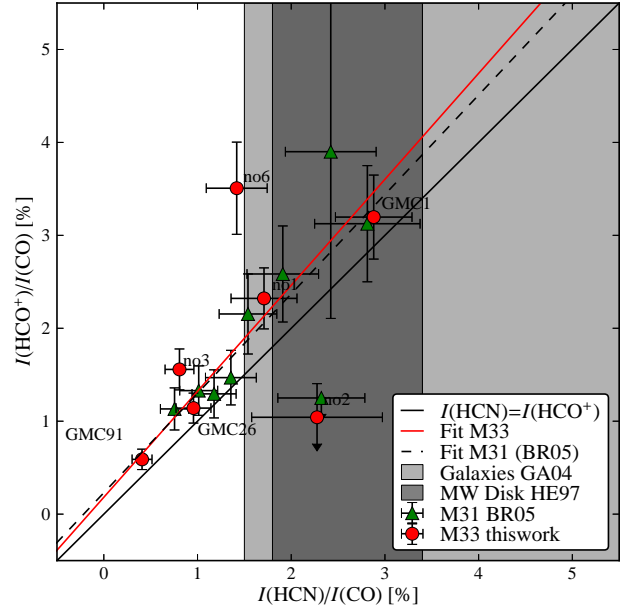}
  \caption{Ratios of integrated intensities HCO$^{+}$/CO vs. HCN/CO for M\,33
    (red points: this work) and for M\,31 (green points:
    BR05). Upper/lower limits are denoted by
    arrows.  Linear least squares fits to data from M\,33 (red solid line) and
    M\,31 (BR05, black dashed) are shown. Both fits exclude points with upper
    limits. The solid black line shows the angle bisector where
    \mbox{$I$(\HCOp)~=~$I$(HCN)}. The gray shaded areas display the range of
    the HCN/CO found in the disk of the Milky Way (MW) by
    \protect\cite[][HE97]{Helfer1997b} (darker gray) and in a sample of normal
    spiral galaxies (GA04b) (lighter gray).}
  \label{Fig:Brouillet}
\end{figure}

For the comparison of different tracers, all data were convolved to the same
resolution of $28''$. We account for the different intrinsic beam sizes of the
CO, HCN, and \mbox{\HCOp\,1--0} observations by multiplying with beam filling
factors determined from the \mbox{\twCO\,2--1} map
\citep{Gratier2010}. Figure~\ref{Fig:Brouillet} compares the \HCOp/CO
vs. HCN/CO line intensities ratios observed in M\,33, with those observed at
nine positions in the disk of the Andromeda galaxy M\,31 \citep[][hereafter
BR05]{Brouillet2005}. M\,31 lies at a similar distance as M\,33 of 780\,kpc and
had been observed with the 30\,m telescope as well. Therefore, both studies
obtain about the same spatial resolution of $\sim114$\,pc.

For M\,33, we find HCN/CO ratios in the range of \mbox{\minHCNCO\,\% --
  \maxHCNCO\,\%} (mean: $\meanHCNCO\pm\stdHCNCO$\,\%) and \HCOp/CO ratios in
\mbox{\minHCOpCO\,\% -- \maxHCOpCO\,\%} (mean:
$\meanHCOpCO\pm\stdHCOpCO$\,\%). BR05 finds comparable values in the spiral
arms of M\,31: HCN/CO \mbox{0.75\,\% -- 2.8\,\%} (mean: $1.7\pm0.5$\,\%) and
\HCOp/CO \mbox{1.1\,\% -- 3.9\,\%} (mean: 2.0$\pm$0.7\,\%). A linear least
squares fit to the M\,33 data results in
\HCOp/CO$~=~($1.14\,$\pm$\,0.15\,\%)\,{\rm
  HCN/CO}\,+\,(0.18\,$\pm$\,0.14\,\%). This is consistent within errors to the
fit results obtained by BR05 for the M\,31 data: \HCOp/CO$~=~$1.07\,\%\,${\rm
  HCN/CO}\,+\,0.23\,\%$. We excluded positions with upper limits from the fit,
i.e.\ position GMC\,no2.

In the Milky Way in the solar neighborhood values of HCN/CO are found between
0.7\,\% -- 1.9\,\% (mean: $1.4\pm2$\,\%), while the Galactic plane hosts on
average $2.6\pm0.8$\,\% \citep{Helfer1997b}. \HCOp/CO values in the Galactic
center range between 0.9\,\% and 7.6\,\% \citep{Riquelme2010}.

The HCN/CO ratios found in the LMC by \cite{Chin1998} and \cite{Heikkilae1999}
lie between \mbox{3\,\% and 6\,\%}, and are thus higher than any value found in
M\,33, M\,31, and also M\,51 where \mbox{HCN/CO~=~1.1\,\% -- 2\,\%} in the spiral
arms are reported \citep{Kuno1995} . Unlike GS04b in a sample of normal
galaxies, we do not find a systematic change in the HCN/CO ratio between
regions in the center of M\,33, i.e. the inner $\sim$\,1\,kpc (here GMC1,
GMC26) and regions at greater galacto-centric distances
(cf. Table\ref{table:values}). GS04b reports that HCN/CO drops from $\sim$ 10\%
in the centers of normal galaxies to \mbox{$\sim$ 1.5\% -- 3\%} in their disks
\mbox{$\gtrapprox$\,4\,kpc}.  In ULIRGs and AGNs the ratios may reach global
HCN/CO values as high as 25\% (GS04b and references therein). GS04b attribute
these high ratios to the presence of starbursts and argue that HCN/CO may serve
as a starburst indicator.

\begin{figure} [t!]
  \centering
  \includegraphics[width=0.48\textwidth]{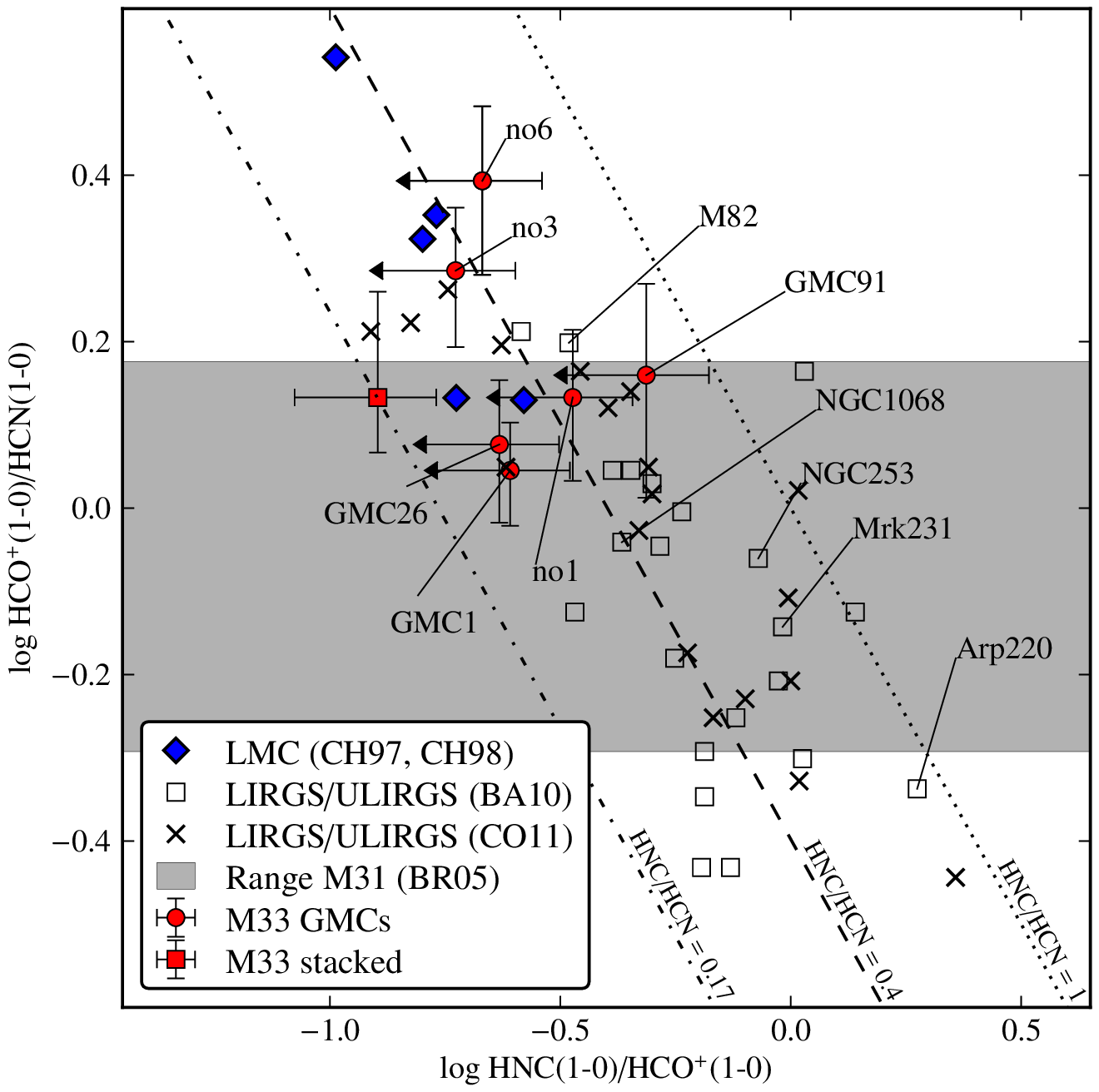}
  \caption{Comparison of integrated intensities \HCOp/HCN vs. HNC/\HCOp in
    M\,33 (red filled circles and square; arrows indicate upper/lower limits)
    with values found in the LMC \protect\citep{Chin1997, Chin1998} (CH97,
    CH98; blue diamonds) and in luminous infrared galaxies compiled by
    \protect\cite{Baan2008} (BA08; open symbols) and by
    \protect\citet{Costagliola2011} (CO11; crosses). The dotted, dashed, and
    dot-dashed lines shows \mbox{HNC/HCN = 1, 0.4, and 0.17} (stacked value of
    M\,33), respectively. The gray shaded area shows the range of the observed
    \HCOp/HCN ratios in M\,31 by BR05.}
  \label{Fig:Baan}
\end{figure}

\subsubsection{\HCOp/HCN vs. HNC/\HCOp}

The \HCOp/HCN ratios observed in M\,33 vary between 1.1 and 2.5, while the
upper limits derived for the HNC/\HCOp\ ratios vary between 0.2 and 0.5
(Fig. \ref{Fig:Baan}). The upper limit of the HNC/HCN ratio is at maximum 0.7
(GMC91), while the stacked spectrum where HNC has been detected shows a HNC/HCN
ratio of 0.17 (Fig.\,\ref{Fig:Baan}).

These ratios are compared with the ratios found in luminous infrared nuclei by
\cite{Baan2008} and \cite{Costagliola2011} and in \HII regions of the LMC by
\citet{Chin1997, Chin1998} (cf. Fig.3a in \citet{Baan2010}). The \HCOp/HCN
ratios are also compared to the range found in the disk of M\,31 by BR05
(cf. Fig.\,\ref{Fig:Baan}).

The \HCOp/HCN ratios, found in M\,33 in the six GMCs with clear detections, lie
at the upper end of the distribution of values found in LIRGs. While the
starburst galaxy M\,82 exhibits a higher ratio than the AGN NGC\,1068 (1.6 vs.
0.9), these ratios lie within the scatter of values and their errors observed
in the disk of M\,33.

The \HCOp/HCN ratios in M\,33 lie in the overlap region between the ones found
in M\,31 and those found in the LMC\@. We find neither ratios as high as in the
LMC, i.e. 3.5, nor ratios as low as in M\,31, i.e. 0.51. Interestingly, all
detected regions in the LMC are situated in the same parameter space as those
detected in M\,33.

The detection of HNC in the stacked spectrum allows us to derive an average
\HCOp/HNC ratio of 7.8 (Table\,\ref{table:stackedTable}) for the GMCs observed
in M\,33. This ratio lies at the very high end of the range of values found in
any of the other samples plotted in Fig.\,\ref{Fig:Baan}.

More remarkably, the HCN/HNC value of 5.9 from the stacked spectra is higher
than any from the other samples we compare with in
Fig.\,\ref{Fig:Baan}. Furthermore, it is higher than ratios observed over the
surface of IC\,342 which are only $\sim1-2$ \citep{Meier2005}, higher than the
ratios in a range of galaxies found by \citet{Huettemeister1995} of
\textless\,4, higher than the typical ratios of 1 observed in starburst and
Seyfert galaxies \citep[e.g.][]{Aalto2002}, and also higher than ratios of 1-3
observed in Galactic molecular complexes \citep{Wootten1978}.

This extraordinary high ratio indicates that the physics or chemistry in M\,33
may be different from that of AGNs and starbursts. The dominance of strong
X-ray radiation in the nuclei of AGNs or even of starbursts may be important
for the differences in the line ratios, since it creates X-ray-dominated
regions (XDRs) that change the chemical abundances
\citep[e.g.][]{Meijerink2005}.

The subsolar metallicity of M\,33 may also play a role in creating such a high
HCN/HNC ratio. However, the HCN/HNC ratio obtained in M\,33 is significantly
higher than those observed in similar low-metallicity environments, such as
N159, 30Dor in the LMC, as well as LIRS36 in the SMC, which are no higher than
3.6 (besides a lower limit of 4.7 in N27 in the SMC) \citep{Chin1998,
  Heikkilae1999} and comparable to the values found in Galactic GMCs
\citep{Huettemeister1995}. Therefore, a subsolar metallicity alone does not
seem to be a guarantee for very high HCN/HNC ratios.

\subsubsection{HCN vs.\ total infrared luminosity ($L_{\rm TIR}$)}
\label{Sect:SFR}

The $L_{\rm TIR}/L_{\rm HCN}$ ratios observed in M\,33
(Table\,\ref{table:values}) range between $1300$ and $3500$
\lsun/K\,kms$^{-1}$\,pc$^2$. These ratios lie at the upper end of the values
found in Milky Way clouds \citep{Wu2010}. Normal galaxies show on average total
$L_{\rm TIR}/L_{\rm HCN}$ ratios of
$\sim$~900$\pm$70\,\lsun/K\,kms$^{-1}$\,pc$^2$
\citep[GS04a,b]{Gracia-Carpio2008}. Higher values, in the range of
\mbox{$\sim$~1100 -- 1700~\lsun/K\,kms$^{-1}$\,pc$^2$}, are found for
\mbox{(U-)LIRGs} \citep[GS04a,b]{Gracia-Carpio2008}, while the highest reported ratios
are found at the extreme end of the LIRG/ULIRG distribution, as well as in
high-z galaxies, and range up to $\sim$~3900 \,\lsun/K\,kms$^{-1}$\,pc$^2$
\citep{Solomon2003, Gao2007, Gracia-Carpio2008, Wu2010,
  Garcia-Burillo2012}. Thus the $L_{\rm TIR}/L_{\rm HCN}$ ratios in M\,33 are
among the highest ratios observed.

In Sect.\,\ref{Sec:PDR} below, we attempt to shed more light on the weak HCN
emission and investigate the line ratios of CO, HCN, and \HCOp\ using models of
photon-dominated regions that take not only the chemical network into account,
but also the detailed heating and cooling processes of a cloud as well as
radiative transfer.

\subsubsection{\twCO/\thCO\ line ratio}

In our sample of GMCs in M\,33 we find \twCO/\thCO line intensity ratios
between 9 and 15. There is no obvious correlation with galacto-centric
distance, FUV strength, or SFR\@. In a study of eight GMCs in the outer disk of
M\,33, \citet{Braine2010b} found similar ratios between 8.9 and 15.7.  In the
Milky Way, the isotopic ratio of $^{12}$C/$^{13}$C varies between values of 80
-- 90 in the solar neighborhood and 20 in the Galactic center
\citep{Wilson1999}. \citet{Polk1988} studied the \twCO/\thCO\ line ratio in the
Milky Way, for several large regions of the plane in comparison with the
average emission from GMCs and of the centers of GMCs. They find that the
average value rises from three in the centers of GMCs, to 4.5 averaged over
GMCs, to 6.7 for the Galactic plane, with peaks of $\sim15$. Their
interpretation is that the higher ratios observed in the plane are caused by
diffuse gas of only moderate optical thickness in $^{12}$CO\@. A similar
interpretation may hold in M\,33. The fraction of dense gas within the beam and
optical depth effects may affect the ratios observed in M\,33.

Ratios in the Magellanic clouds are observed by \cite{Heikkilae1999} to cover
values between 5 and 18, a somewhat wider range than found in M\,33. Unlike in
M\,33, a gradient is seen on larger scales in a set of IR-bright nearby galaxies,
dropping from values of about ten in the center to values as low as two at
larger radii \citep{Tan2011}. Although \citet{Aalto1995} finds variations with
galacto-centric radius in some galaxies of their IR-bright sample, other
galaxies exist where the \twCO/\thCO stays constant with radius. They report a
mean value of $\sim$\,12 for the centers of most galaxies in their sample
except for the most luminous mergers with ratios of $\gtrapprox$\,20 \citep[see
also][]{Casoli1992}.

\section{Molecular abundances}
\label{Sect:ColDens}

\begin{table}
  \caption{Molecular abundances in M\,33 and
    typical examples of galactic and extra-galactic sources.}
  \label{table:abundances}
  \centering
  \begin{tabular}{lccccl}
    \hline \hline
    Source      & \HCOp & HCN    & HNC     & C$_2$H & References \\
    \hline
    M\,33 min & $-$10.3 & $-$9.8 & - & - & \\
    M\,33 stacked & $-$9.8 & $-$9.2 & $-$10.5 & $-$8.4 & \\
    M\,33 max & $-$9.5 & $-$9.1 & - & - & \\
    LMC N159    & $-9.7$  & $-9.7$ & $-10.2$ & -      & 1,2 \\
    \hline
    NGC\,253    & $-8.8$  & $-8.3$ & $-9.0$  & $-7.7$ & 1,3 \\
    M\,82         & $-8.4$  & $-8.4$ & $-8.8$  & $-7.6$ & 1,3 \\
    IC\,342 GMC-A &       &        & $-8.7$  & $-7.5$ & 5 \\
    \hline
    Orion Bar   & $-8.5$  & $-8.3$ & $-9.0$  & $-8.7$ & 1 \\
    TMC-1       & $-8.4$  & $-7.7$ & $-7.7$  & $-7.1$ & 1,6 \\
    Transl. Cl. & $-8.7$  & $-7.4$ & $-8.6$  & -      & 1,4 \\
    \hline
  \end{tabular}
  \tablefoot{Entries show  $\log(N(X)/N({\rm H}_2))$.}
  \tablebib{(1)~\protect\citet{Omont2007}; (2)~\citet{Johansson1994};
    (3)~\protect\citet{Martin2006a}; (4)~\protect\citet{Turner2000};
    (5)~\protect\citet[][]{Meier2005} and references therein;
    (6)~\protect\citet{HogerheijdeSandell2000}}
\end{table}

We use the observed line intensities of \HCOp, HCN, HNC, and C$_2$H to estimate
the molecular column densities and abundances, assuming local thermodynamic
equilibrium (LTE) and optically thin emission. Details on the calculations are given
Appendix~\ref{Appendix:LTE}, with results shown in Table~\ref{Tab:Coldens}. The
abundances are only lower limits in case the emission of HCN and \HCOp\ is
optically thick. In comparison with our PDR-model analysis below we find,
however, that for the best-fitting models to our observations the optical
depths ($\tau$) in the centers of the lines of HCN and \HCOp are
$\tau$~\textless~0.1, suggesting that emission is likely to be optically thin
(cf. Sect. \ref{Section:Comparison}).  In Table\,\ref{table:abundances}, we
compare the abundances with those found in other galaxies (LMC, NGC\,253, M82,
IC\,342), in selected Galactic sources, the photon-dominated region Orion Bar,
the dark cloud \mbox{TMC-1}, and a translucent cloud. The estimated column
densities for the stacked values are given in Table~\ref{table:stackedTable}.

The abundances of \HCOp, HCN, and HNC found in M\,33 are very similar to those
found in the LMC cloud N159. The abundances derived from the stacked spectrum
of M\,33 agree to within 0.5\,dex with those of N159. Galactic sources have
values that are higher by more than an order of magnitude. The Orion Bar, for
example, shows 1.8\,dex to 0.8\,dex higher abundances. The good agreement with
the LMC may be driven by its similar metallicity of \mbox{0.3--0.5} relative to
the solar metallicity \citep{Hunter2007}, which is only slightly lower on
average than in M\,33 \citep{Magrini2007, Magrini2010}. However, the
  C$_2$H abundance observed in M\,33 and in the Orion Bar agree within 0.3\,dex.

The LTE \HCOp/HCN abundance ratios in M\,33 range between 0.2 and 0.5
(Table\,\ref{Tab:Coldens}). \citet{Godard2010} measured an \HCOp/HCN abundance
ratio of 0.5 in the diffuse ISM of the Milky Way, similar to the ratio found in
the solar neighborhood, and similar to the higher ratios found in M\,33.

%%%%%%%%%%%%%%%%%%%%%%%%%%%%%%%%%%%%%%%%%%%%%%%%%%%%%%%%%%%%%%%%%%%%%%%%%%%%%%

\section{PDR models}
\label{Sec:PDR}

\subsection{Setup}

To improve on the LTE analysis and to better understand why HCN is less
luminous than \HCOp\ in M\,33, we compare the observed \HCOp/HCN, HCN/CO, and
\HCOp/CO line ratios with models of photon-dominated regions (PDRs) using the
Meudon PDR code \citep{LePetit2006, Gonzalez2008}. The line intensities of the
molecules \thCO, HNC, and C$_2$H are not modeled.

We ran a grid of models for different densities \nh~=~0.1, 0.5, 1, 5, 10, 50,
10$^2$ $\times$ 10$^4$\,cm$^{-3}$, FUV fields \Gn$~=~10, 50, 100$ in Habing
units\footnote{Habing units correspond to an average interstellar radiation
  field (ISRF) between 6\,eV $\le h\nu \le$ 13.6\,eV of
  1.6\,10$^{-3}$erg\,cm$^{-2}$\,s$^{-1}$ \citep{Habing1968}. Another unit that
  is frequently used is the Draine unit of the local average ISRF, which is 1.7
  $\times$ \Gn in Habing units.}, and optical extinctions \mbox{\Av~=~2 --
  50\,mag}, i.e.\ in steps of \mbox{log\,\Av$\sim$0.2}. We calculated this grid
of models for a solar and a subsolar metallicity. The subsolar one is tailored
to describe the metallicity in the disk of M\,33. See Appendix \ref{PDRsetup}
for a detailed description of the model setup.

\subsection{Results}

\begin{figure*}[ht!]
  \centering
  \label{ratioModels}
  \subfloat{\label{Fig:hcoHcn} \includegraphics[width=0.99\textwidth, angle=0]
    {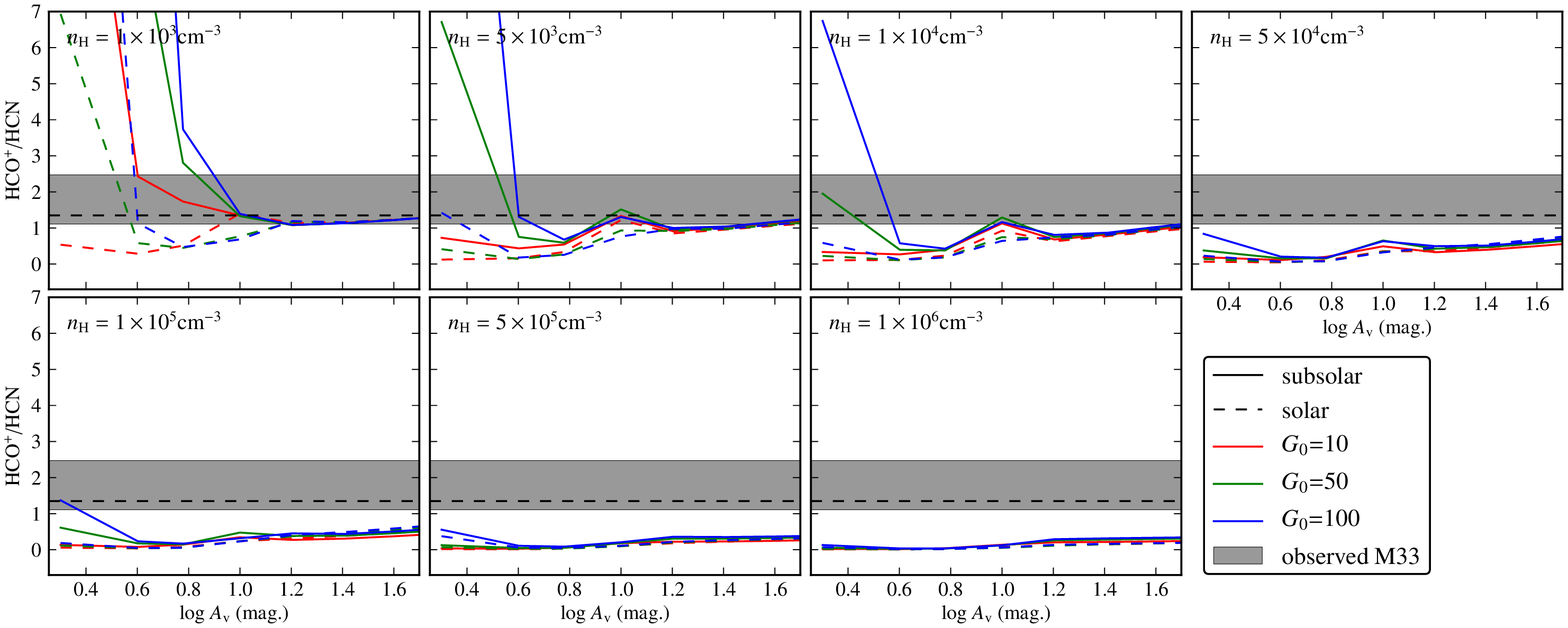}} \vskip0.01\textwidth
  \subfloat{\label{Fig:hcnCo} \includegraphics[width=0.99\textwidth, angle=0]
    {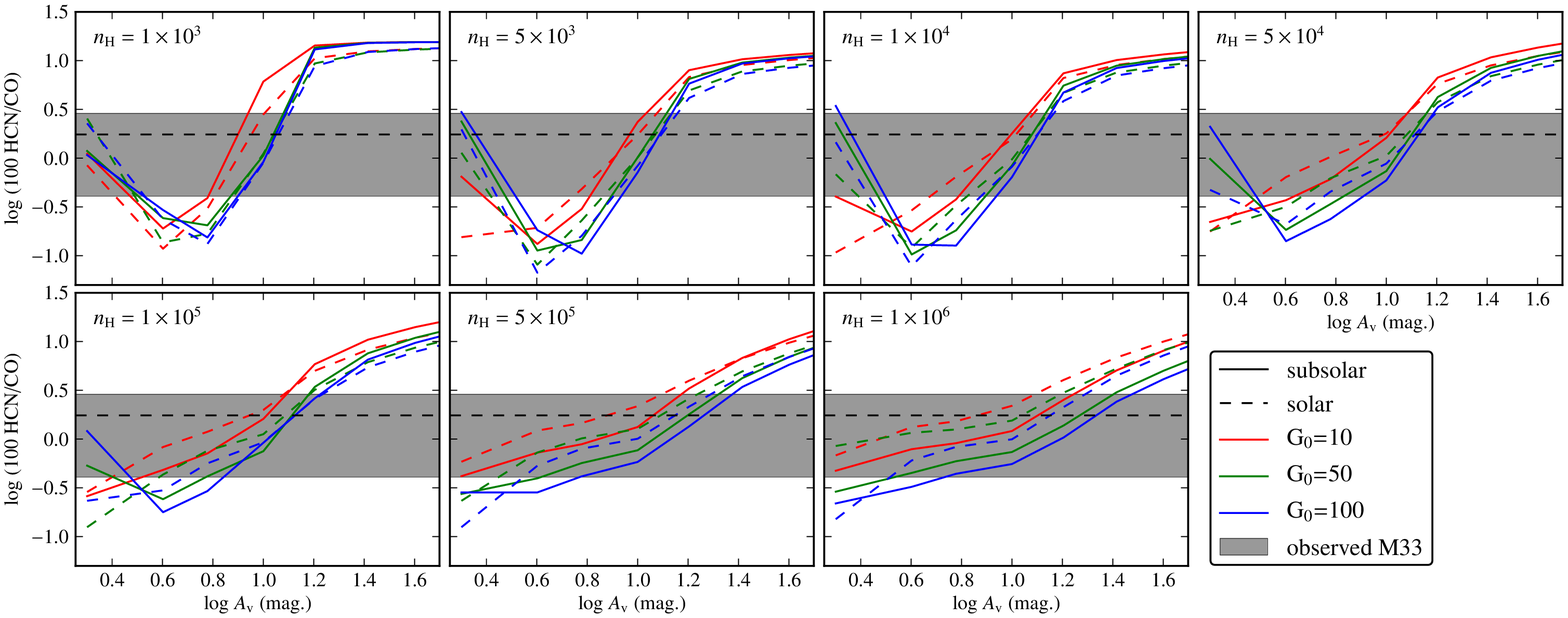}}
  \caption{PDR model line ratios for subsolar (solid lines) and solar
    metallicity (dashed lines): \HCOp/HCN (top) and HCN/CO (bottom).
    Different colors indicate different FUV field strengths \Gn = 10 (red), 50
    (green), and 100 (blue). Every panel of a subfigure shows the results for
    one density; from left to right and top to bottom \nh~=~0.1, 0.5, 1, 5, 10,
    50, and 10$^2$ $\times$ 10$^4$\,cm$^{-3}$. Gray areas mark the range of
    observed ratios in M\,33. The dashed horizontal lines show the values from
    the stacked spectra.}
  \label{Fig:modelResults}
\end{figure*}

The modeled HCN/CO line ratio (Fig.\,\ref{Fig:modelResults},
Tables\,\ref{table:PDRResultsSubSolar} and \ref{table:PDRResultsSolar}) hardly
varies with metallicity, \nh or FUV. It varies, however strongly, with
\Av. After an initial drop of upto an order of magnitude for extinctions less
than about 4 mag, it rises until \Av $\sim$20\,mag and then saturates at a
value that is nearly independent of any of the input parameters.

In contrast, the modeled \HCOp/HCN ratio shows differences between the two
metallicities for \Av$\lesssim$10\,mag. The subsolar model shows higher ratios
than the solar model for a given \Av, FUV field, and density. Strong FUV fields
and low densities increase the ratio. At higher optical extinctions the ratios
are hardly influenced by changes in metallicity, \Av or FUV\@. This reflects
the variation in HCN and \HCOp\ abundances. At low \Av, subsolar HCN abundances
are lower than solar ones by up to a factor of 0.6\,dex, while \HCOp\ is
enhanced in the subsolar models by up to 0.7\,dex. Also the CO abundances show
a clear dependence on metallicity. For all input parameters, its abundance in
the subsolar models is $\sim$0.6\,dex lower than in the solar models. This
directly reflects the underabundance of carbon of 0.6\,dex in the subsolar
models. As a result, the HCN/CO line ratio is fairly independent of
metallicity.

For optical extinctions in the range of \Av$\leq$\,8 mag, where the bulk of
molecular gas in galaxies resides \citep{Tielens2005}, the HCN/CO ratio
increases with increasing densities. In general in LIRGs/ULIRGs where most of
gas has higher density than in normal galaxies, the HCN/CO is also higher
(GS04a,b).

\begin{table*}[t!]
  \center
  \caption{Best-fitting PDR models}
  \label{modelsSources}
  \begin{tabular}{ccccccccc}
    \hline
    \hline

    & \HCOp/HCN	& HCN/CO   & \Av      & \nh             & FUV       &
    $\Phi_{A_{\rm v}}$\tablefootmark{a} & $\Phi_{\rm FUV}$\tablefootmark{b}
    &best $\chi^2$ \tablefootmark{c}\\
    &       & [\%]       &  [mag.]  &     [cm$^{-3}$] &     [\Gn] &              &            &             \\
    \hline
    \hline
    \multicolumn{9}{c}{subsolar metallicity Models} \\
    \hline
    Stacked	&	1.4 $\pm$ 0.2	&	1.0 $\pm$ 0.1	&	8 $\pm$ 3	&	(3 $\pm$ 4) 10$^4$	&	68 $\pm$ 24	&	0.5 $\pm$ 0.2	&	0.3 $\pm$ 0.1	&	0.1 $\pm$ 0.1	\\
    NO6	&	2.5 $\pm$ 0.2	&	1.4 $\pm$ 0.3	&	4 $\pm$ 2	&	(6 $\pm$ 4) 10$^3$	&	41 $\pm$ 16	&	1.7 $\pm$ 1.1	&	0.9 $\pm$ 0.4	&	1.4 $\pm$ 0.5	\\
    NO3	&	1.9 $\pm$ 0.1	&	0.8 $\pm$ 0.2	&	8 $\pm$ 2	&	(3 $\pm$ 2) 10$^3$	&	27 $\pm$ 20	&	0.7 $\pm$ 0.2	&	0.8 $\pm$ 0.6	&	0.6 $\pm$ 0.3	\\
    GMC91	&	1.4 $\pm$ 0.3	&	0.4 $\pm$ 0.1	&	6 $\pm$ 2	&	(1 $\pm$ 5) 10$^4$	&	54 $\pm$ 44	&	0.9 $\pm$ 0.3	&	0.2 $\pm$ 0.2	&	0.1 $\pm$ 0.1	\\
    NO1	&	1.4 $\pm$ 0.2	&	1.7 $\pm$ 0.4	&	7 $\pm$ 4	&	(3 $\pm$ 4) 10$^4$	&	42 $\pm$ 40	&	0.3 $\pm$ 0.2	&	0.3 $\pm$ 0.3	&	0.2 $\pm$ 0.1	\\
    GMC1	&	1.1 $\pm$ 0.1	&	2.9 $\pm$ 0.4	&	10 $\pm$ 3	&	(1 $\pm$ 2) 10$^4$	&	30 $\pm$ 37	&	0.6 $\pm$ 0.2	&	1.7 $\pm$ 2.1	&	0.7 $\pm$ 0.5	\\
    GMC26	&	1.2 $\pm$ 0.1	&	1.0 $\pm$ 0.2	&	9 $\pm$ 2	&	(1 $\pm$ 3) 10$^4$	&	67 $\pm$ 24	&	0.3 $\pm$ 0.1	&	0.2 $\pm$ 0.1	&	0.1 $\pm$ 0.1	\\
    \hline
    \hline
    \multicolumn{9}{c}{solar metallicity models} \\
    \hline
    Stacked	&	1.4 $\pm$ 0.2	&	1.0 $\pm$ 0.1	&	10 $\pm$ 1	&	(5 $\pm$ 1) 10$^3$	&	47 $\pm$ 12	&	0.4 $\pm$ 0.0	&	0.5 $\pm$ 0.1	&	0.6 $\pm$ 0.4	\\
    NO6	&	2.5 $\pm$ 0.2	&	1.4 $\pm$ 0.3	&	2 $\pm$ 0	&	(5 $\pm$ 1) 10$^3$	&	100 $\pm$ 4	&	3.1 $\pm$ 0.8	&	0.4 $\pm$ 0.0	&	3.4 $\pm$ 1.2	\\
    NO3	&	1.9 $\pm$ 0.1	&	0.8 $\pm$ 0.2	&	6 $\pm$ 3	&	(2 $\pm$ 2) 10$^3$	&	81 $\pm$ 32	&	0.9 $\pm$ 0.5	&	0.3 $\pm$ 0.1	&	3.4 $\pm$ 0.9	\\
    GMC91	&	1.4 $\pm$ 0.3	&	0.4 $\pm$ 0.1	&	4 $\pm$ 1	&	(1 $\pm$ 1) 10$^3$	&	98 $\pm$ 11	&	1.4 $\pm$ 0.3	&	0.1 $\pm$ 0.0	&	0.5 $\pm$ 0.4	\\
    NO1	&	1.4 $\pm$ 0.2	&	1.7 $\pm$ 0.4	&	7 $\pm$ 4	&	(5 $\pm$ 3) 10$^3$	&	50 $\pm$ 43	&	0.4 $\pm$ 0.2	&	0.3 $\pm$ 0.2	&	0.2 $\pm$ 0.2	\\
    GMC1	&	1.1 $\pm$ 0.1	&	2.9 $\pm$ 0.4	&	11 $\pm$ 2	&	(3 $\pm$ 6) 10$^3$	&	26 $\pm$ 34	&	0.6 $\pm$ 0.1	&	1.9 $\pm$ 2.6	&	0.5 $\pm$ 0.2	\\
    GMC26	&	1.2 $\pm$ 0.1	&	1.0 $\pm$ 0.2	&	9 $\pm$ 2	&	(5 $\pm$ 1) 10$^3$	&	52 $\pm$ 17	&	0.2 $\pm$ 0.1	&	0.2 $\pm$ 0.1	&	0.4 $\pm$ 0.3	\\
    \hline

  \end{tabular}

  \tablefoot{\tablefoottext{a}{beam-filling factor derived from \Av{}(\twCO)/\Av{}(Model);}
    \tablefoottext{b}{beam-filling factor derived from FUV{}(TIR)/FUV{}(Model);}
    \tablefoottext{c}{average $\chi^2$ of the best-fitting models.}}

\end{table*}

\section{{Comparison with PDR models}}
\label{Section:Comparison}

Figure\,\ref{Fig:modelResults} shows the range in observed intensity ratios
(cf.\,Table\,\ref{table:values}), where the plot shows that the optical
extinctions play a decisive role in determining the line ratios. High
extinctions of \Av\textgreater 16\,mag are inconsistent with the observed
HCN/CO ratios. Similarly, high densities of 10$^5$\,cm$^{-3}$ cannot reproduce
the measured \HCOp/HCN ratios. Interestingly, metallicity only plays a minor
role. In general, both metallicity models allow the observed range of line
ratios to be reproduced.

To quantify the agreement between the line ratios of the different models and
the individual observed clouds including the stacked spectrum, we use a
$\chi^2$ fit routine. To get a handle onto the errors of the best-fitting
models, a Monte Carlo analysis is employed. The details on the fitting and
Monte Carlo methods are given in Appendix~\ref{chisquare}.
Table~\ref{modelsSources} shows the input parameters \Av, \nh, and FUV of the
best-fitting subsolar and solar metallicity models, i.e.\ those having the
lowest $\chi^2$, for the stacked values and each observed cloud.

\subsection{Stacked ratios}

The best-fitting models for reproducing the stacked \HCOp/HCN and HCN/CO ratios
of 1.4 and 1\%, respectively, that describe the averaged GMC properties are
\mbox{\Av\,=\,8\,mag}, \mbox{\nh\,=\,3\,10$^4$\cmt}, and
\mbox{FUV\,=\,68\,\Gn}. Emission stems from moderately dense gas with average
line-of-sight column densities of 8\,mag. The beam-filling factor $\Phi_{\rm
  FUV}$ derived from ratio of the beam-averaged TIR intensity to the fitted
local FUV field is $\sim$\,30\%; i.e., the fitted FUV field strengths are
significantly higher than expected from the observations. The same holds for
the beam filling factors deduced from the ratios of extinctions derived from CO
and the \Av of the best-fitting models, which are about $\Phi_{A_{\rm
    v}}$\,=\,50\%. This is not surprising, however, and indicates that emission
is clumped within the 114\,pc beam. From the models of the calculated grid
closest to the best fit, i.e., Av~=~6 and 10\,mag, \nh~=~1\,10$^{4}$, and
FUV~=~50\,\Gn, we find optical depths ($\tau$) in the centers of the lines of
HCN and \HCOp that lie between \mbox{0.02 -- 0.1} and \mbox{0.07 -- 0.1},
respectively. Thus both lines are optically thin. \twCO is moderately optically
thick with optical depths of $\tau$~4~--~25. The line width assumed in the
Meudon PDR code is $\sim$~3\kms (cf. Appendix\,\ref{PDRsetup}).

\subsection{Individual regions}

Here, we focus on individual regions grouped by their particular \HCOp/HCN
ratios and thus their best-fit \Av values: no6 showing a high ratio of 2.5,
GMC91, no3, and no1 have intermediate ratios of 1.4-1.9 and GMC26, GMC1 having
ratios of 1.1-1.2.

\paragraph{GMC\,no6}
This cloud shows the highest \HCOp/HCN ratio of 2.5, while at the same time the
HCN/CO ratio is  relatively low with 1.4 and weaker than expected from the
linear fit to the M\,33 data (cf. Sect. \ref{co-ratios} and
Fig. \ref{Fig:Brouillet}). GMC\,no6 is best fitted by subsolar models that
yield a low best-fitting value for \Av of $\sim$ 4 mag, while the best-fitting
density and FUV strength are 6\,10$^3$\,cm$^{-3}$ and 40\,\Gn,
respectively. The beam-filling factor derived from \Av is 1.7, indicating that
emission completely fills the beam with several clouds along the
line-of-sight. This cloud has the second highest star formation rate of our
sample of 35.9\,M$_{\sun}$\,Gyr$^{-1}$\,p$c^{-2}$, and the same holds for the FUV
field strength of 37.3\,G$_0$.

\paragraph{GMC91, no3, and no1}

The line ratios of these three clouds are best described by subsolar
models. The best-fitting \Av are similar with 6--8\,mag. So are the FUV
\mbox{30--50\,\Gn} and the densities 3\,10$^3$\,cm$^{-3}$ --
3\,10$^4$\,cm$^{-3}$, and no1 and no3 have similar SFR rates of
$\sim$~13\,M$_{\sun}$\,Gyr$^{-1}$\,p$c^{-2}$, while no3 is a factor of four more
massive than no1 with \mbox{$M_{\rm H_2}$~=~8\,10 $^5$ M$_{\sun}$}. GMC91 lies
at only 320\,pc distance in close vicinity of GMC\,no3 and is only slightly
more massive than the same. It is the most CO intense cloud in our sample
while its HCN and \HCOp emission is relatively weak. This renders GMC91
somewhat peculiar and results in a low \HCOp/CO ratio of 0.6\,\% and, as
already found by RPG11, a particularly low HCN/CO intensity ratio of 0.4\,\%
(Table\,\ref{table:values}). The HCN/CO ratio of GMC91 is much lower than the
ratios observed in the disk of the Milky Way by \citet{Helfer1997b}, who find
\mbox{2.6\,\%\,$\pm$\,0.8\,\%} and also in the inner disks (5--10\,kpc) of
normal galaxies by GS04b who find \mbox{4\,\%\,$\pm$\,2\,\%}. The relatively
weak HCN and \HCOp emission may indicate a low fraction of dense mass in GMC91,
which thus may be a rather quiescent GMC with only a low SFR\@. Indeed, it has
the lowest star formation rate of 4\,M$_{\sun}$\,Gyr$^{-1}$\,p$c^{-2}$ of all
observed clouds.

\paragraph{GMC1 and GMC26}

These two clouds host the lowest observed ratios of \HCOp/HCN of 1.1 -- 1.2.
Here, the solar models provide slightly better or equal fits than the subsolar
models. However, since the ISM of M\,33 is subsolar, here we discuss only the
best fits to the subsolar models. These GMCs have similar best-fitting input parameters of
\Av~9--10\,mag and \mbox{\nh~$\sim$~10$^4$\,cm$^{-3}$}, while the best-fitting
FUV field strengths are \mbox{$\sim$ 30\,\Gn} and 70\,\Gn, respectively. However
comparing their physical properties in Table~\ref{table:values}, again, these
two clouds are actually not at all alike. GMC1 is located in the very center of
M\,33 and is by far the most massive cloud in our sample. It is actively forming
stars at a rate of 65 M$_{\sun}$ Gyr$^{-1}$ p$c^{-2}$, the highest in our
sample, and has an correspondingly high FUV field of 50.7\,\Gn. GMC26 has much
lower HCN/CO ratios, and its SFR is only 6.6\,M$_{\sun}$\,Gyr$^{-1}$\,p$c^{-2}$
the second lowest in the sample with exception of no2.

For all best-fitting solutions of the six individual positions we find that the
modeled optical depths of HCN and \HCOp are $\tau$~$\leq$~0.1, which renders
emission of these lines to be likely optically thin. This also justifies the
assumption of optical thin emission for the LTE analysis in
Sect\,\ref{Sect:ColDens}. Indeed, the PDR modeled abundances of both molecules
are comparable to the ones derived from LTE (cf. Table~\ref{table:abundances}
and \ref{table:PDRResultsSolar}).

In conclusion, it is noteworthy to repeat that the line ratios studied here are
fairly independent of the metallicity, SF activity, and FUV field strength of
the parent GMC, while the optical extinction has a major influence on the
modeled line ratios.

\section{Summary}

We present IRAM 30m observations of the ground-state transitions of HCN, \HCOp,
\twCO\, and \thCO\ of seven GMCs distributed along the major axis in the disk of
the nearby spiral galaxy M\,33. We achieve a spatial resolution of
$\sim$\,114\,pc at a frequency of 89\,GHz.

The molecular gas masses of the target GMCs vary by a factor of $\sim$\,130
between 0.1\,10$^5$\,\msun\ (GMC no2) and 13\,10$^5$\,\msun\ (GMC1) and the
star formation rates derived from H$\alpha$ and 24\,\mum images vary by a
factor of more than 50. The FUV field strengths show a variation of more than a
factor 20. Below, we summarize the main results.

\begin{enumerate}

\item For the six GMCs where \HCOp\ is detected, peak line temperatures (on the
  $T_{\rm mb}$ scale) vary between 6 and 12\,mK. The \HCOp/HCN-integrated
  intensity line ratios lie between 1.1 and 2.5 (on the K\,km\,s$^{-1}$ scale,
  cf. Table~\ref{table:values}). Similar line ratios are observed in the disk
  of M\,31 \citep{Brouillet2005}.

\item The line intensity ratios HCN/CO and \HCOp/CO vary between
  \mbox{$(0.4-2.9)\,\%$} and \mbox{$(0.6-3.5)\,\%$}, respectively. The spread
  of ratios found in M\,33 is slightly larger than in the spiral arms of M\,31
  \citep[][Fig\,\ref{Fig:Brouillet}]{Brouillet2005}. GMC\,91 exhibits a
  particularly low HCN/CO ratio of 0.4\,\%, which is much lower than values in
  the Galactic disk of \mbox{2.5\,\%\,$\pm$\,0.6\,\%} \citep{Helfer1997b} or in
  normal galaxies with \mbox{4\,\%\,$\pm$\,2\,\%} (GS04a).

\item The $L_{\rm TIR}/L^\prime_{\rm HCN}$ luminosity ratios range
  between $1.3\,10^3$ and $3.5\,10^3$ and are situated at the
  very high end of ratios found by \citet{Wu2010} in molecular clouds of the
  Milky Way and LIRGs/ULIRGs. This shows that HCN emission in comparison to the
  $L_{\rm TIR}$ in M\,33 particularly weak.

\item Stacking of all spectra taken at the seven GMC positions leads to
  $3\,\sigma$ detections of CCH and HNC\@. The HCN/HNC ratio of 5.8 is
  remarkable high. It is higher than values found in the LMC \citep{Chin1997,
    Chin1998}, in IC\,342 \citep{Meier2005}, in samples of LIRGs/ULIRGs
  \citep{Baan2008, Costagliola2011}, in starburst and Seyfert galaxies
  \citep[e.g.][]{Aalto2002}, and in Galactic molecular complexes
  \citet{Wootten1978}, where all together no values higher than three are
  reported.

\item The \HCOp, HCN, HNC abundances, derived assuming LTE\@, agree with those
  of the LMC cloud N159 within 0.5\,dex. In contrast, the Orion Bar, a Galactic
  massive star-forming region, shows significantly higher abundances of all
  three tracers by 0.8\,dex to 1.8\,dex. These striking differences may reflect
  the factor two subsolar metallicities of both the LMC and M\,33.

\item Employing the Meudon PDR code to model photon-dominated regions we
  investigated the influence of the metallicity on the abundances and emission
  of HCN and \HCOp. For a range of optical extinctions, volume densities, and
  FUV radiation field strengths, we derived two sets of models with different
  metallicity, one reflecting the abundances in the Orion nebula by
  \citet{Simon-Diaz2011a}, the other the average subsolar metallicity of M\,33
  \citep{Magrini2010}.

  Both sets of models are able to describe the observed range of \HCOp/HCN and
  HCN/CO line ratios reasonably well ($\chi^2<3.4$). Therefore, changes in
  metallicity do not need to be invoked to describe the observed line
  ratios. The observations are described by subsolar models with optical
  extinctions between 4\,mag and 10\,mag and moderate densities of \textless
  3\,10$^4$\,\cmt, with little influence by the FUV field strength. The optical
  extinction has a pronounced influence on the modeled ratios, while FUV field,
  metallicity and even density only play minor roles. The modeled lines of HCN
  and \HCOp of the best-fitting models are found to be optically thin with
  optical depths $\tau$~$\leq$\,0.1.

\end{enumerate}

{\tiny
  \paragraph*{Acknowledgments} PG is supported by the French Agence Nationale
  de la Recherche grant ANR-09-BLAN-0231-01 as part of the SCHISM project.}

%%
% Add the Bibliography
\bibliographystyle{aa} \bibliography{denseGasInM33-publisher}

\Online

\begin{appendix}

\section{PDR Model setup}
\label{PDRsetup}
\begin{table}
  \caption{Initial Abundances used in the PDR models}
  \centering{}
  \begin{tabular}{ccccc}
    \hline
    \hline
    Species & Sun\tablefootmark{a} & Orion\tablefootmark{b} & M33\tablefootmark{c} & diff.\ in [dex]\tablefootmark{d}\\
    \hline
    H$_2$  & 0.1 & 0.1 & 0.1 & \\
    H     & 0.8 & 0.8 & 0.8 & \\
    He    & 0.1 & 0.1 & 0.1 & \\
    C    & 8.43 & 8.37 & 7.77 & 0.60 \\
    O    & 8.69 & 8.65 & 8.27 & 0.38 \\
    N    & 7.83 & 7.92 & 7.31 & 0.61 \\
    S    & 7.12 & 6.87 & 6.71 & 0.16 \\
    Fe    & 7.50 & 6.00 & 5.73 & 0.27 \\
    \hline
  \end{tabular}
  \label{tab:metal}
  \tablefoot{Entries show 12+log($n$(X)/(2$n$(H$_2$)+$n$(H)).
    \tablefoottext{a}{\protect\citet{Asplund2009};}
    \tablefoottext{b}{\protect\citet{Simon-Diaz2011a};}
    \tablefoottext{c}{\protect\citet{Magrini2010, Henry2000};}
    \tablefoottext{d}{between Orion and M\,33}.
    }
\end{table}

We use the Meudon PDR code \citep{LePetit2006, Gonzalez2008}, which solves the
thermal balance, chemical network, detailed balance, and radiative transfer of
a plane-parallel slab of optical extinction \Av and constant density
\nh illuminated from both sides by an FUV field of intensity \Gn. The modeled
line widths are calculated by the Meudon code via the Doppler broadening of the
gas due to the kinetic temperature and the turbulent velocity of the gas. The
latter dominates the line widths and is an input parameter to the Meudon code.
We use the default value of the Meudon code which, expressed
in terms of the FWHM of the velocity distribution, is ($\Delta\,v$)$_{\rm
  FWHM}$~$\sim$~3\,\kms. This is slightly lower than the observed line widths
that are in the range of 4--11\,\kms.  We also used the defaults for additional
input parameters (e.g.\ the cosmic-ray ionization rate and size distribution of
dust grains).

The grid of models has been calculated for volume densities \nh~=~0.1, 0.5, 1, 5, 10,
50, 10$^2$ $\times$ 10$^4$\,cm$^{-3}$, FUV field strengths \Gn$~=~10, 50, 100$ in Habing units, and
optical extinctions \mbox{\Av~=~2 -- 50\,mag}, i.e.\ in steps of
\mbox{log\,\Av$\sim$0.2}.

The input range of FUV fields is motivated by the range of FUV fields derived
from the TIR in the six observed clouds which vary between \Gn$=11.3$ and 50.7,
excluding position no2 with a very low \Gn of 2.5
(cf. Table~\ref{table:values}). We add models of \mbox{\Gn=100} to cover
possible higher local radiation fields. The values of $A_{\rm V}$ range from
those found in translucent clouds with 2\,mag to extinctions of 50\,mag
typically found in resolved Galactic high-mass star-forming clouds. The
densities cover values found in a typical molecular cloud, covering the
critical densities of the observed tracers and transitions.

We calculated this grid of models for solar and subsolar
metallicities. Table\,\ref{tab:metal} shows the abundances measured in M\,33 and
in the Orion nebula which we use as initial abundances for our PDR models. The
abundances of the sun are listed for comparison. For the ``solar'' metallicity
model, we adopt the abundances of O, C, N, S, Fe
measured in the \HII region of the Orion nebula by \citet{Simon-Diaz2011a}. The
heavy elements S and Fe are depleted to dust grains with respect to the
abundances found in the solar photosphere.

For the subsolar metallicity PDR model, we adopt the averaged O, N, and S
abundances measured in M\,33 from \citet[][]{Magrini2010} who targeted 33 \HII
regions between 1 and 8\,kpc galacto-centric distance. The carbon gas-phase
abundance has not been measured in M\,33. We estimate it using the metallicity
dependence of the C/O ratio described by \citep{Henry2000}. To derive the Fe
abundance, we scale the solar Fe/O ratio by the subsolar O metallicity of M\,33.

\end{appendix}

\begin{appendix}

\begin{figure*} [ht!]
  \centering
  \includegraphics[width=0.8\textwidth]{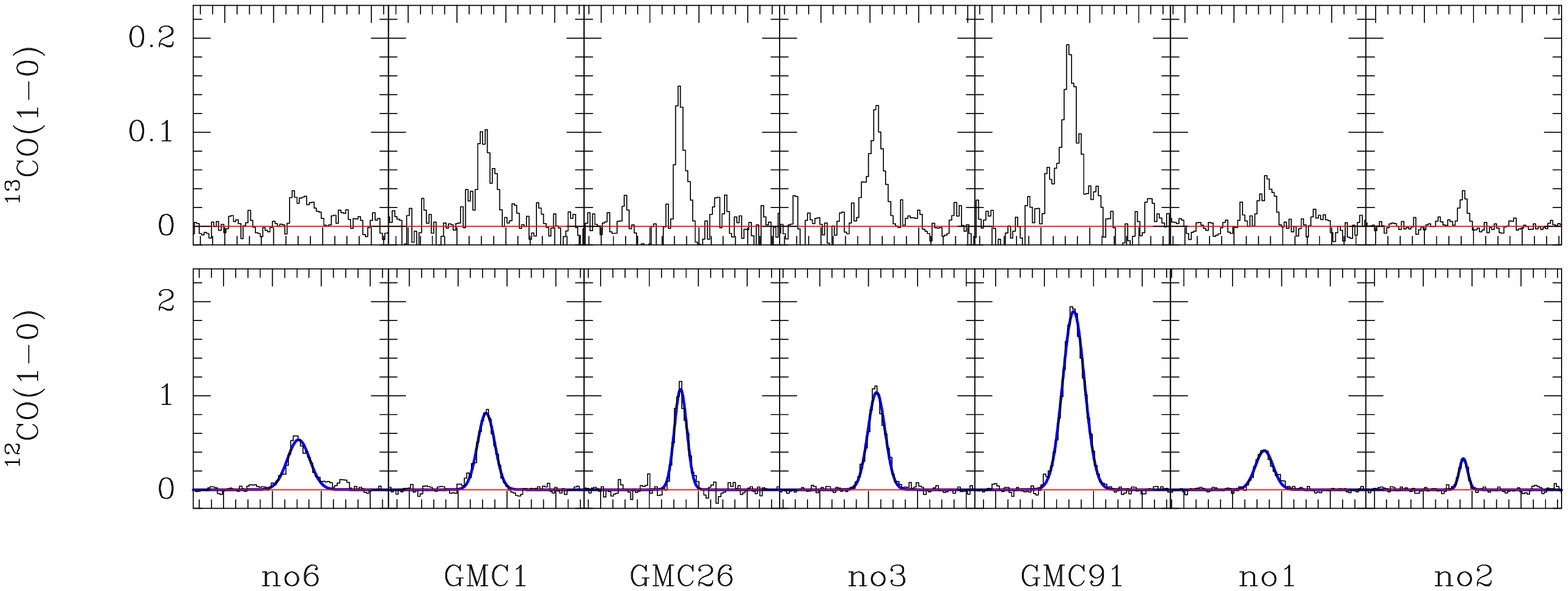}
  \caption{\twCO and \thCO 1-0 spectra at the seven selected GMC positions at
    high-velocity resolution of 1\,\kms. The \twCO spectra of positions
    GMC91, GMC26 and GMC1 are taken from RPG11. The blue lines overplotted on
  the \twCO spectra show the Gaussian line fits used to deduce the FWHM of the
  lines which are listed in Table~\ref{table:values}.}
  \label{Fig:COFull}
\end{figure*}

\section{Chi-squared fitting and Monte Carlo analysis}
\label{chisquare}

To quantitatively compare the modeled and observed line ratios we calculate the
summed squared residuals weighted by the observational errors between the
modeled and observed line ratios \HCOp/HCN and HCN/CO, for every model and for
each observed cloud as

\begin{equation}
  \label{eq:redChi}
  \chi^2 = \frac{1}{2} \sum_{i=1}^2 \frac{({\rm Ratio_{obs, i}} - {\rm
      Ratio_{mod, i}})^2}{\sigma_i^2} ,
  \label{residual}
\end{equation}

\noindent where $\sigma$ is the observational error of the line ratios deduced from the
errors of the line intensities, cf. Sect.~\ref{observations}. We normalize the
$\chi^2$ by dividing by $N=2$, the number of independent observed line ratios.
The minimum $\chi^2$ gives the best-fitting model.

To get a handle on the errors of the best-fitting models, we employ a Monte
Carlo simulation. We assume that the measured ratios and their errors follow a
Gaussian distribution with an expected value equal to the ratio and a variance
equal to $\sigma^2$.

We generate 5000 sets of HCN/CO and HCO+/HCN ratios drawn randomly from their
Gaussian parent distribution. For every set we calculate a $\chi^2$ as
explained above and deduce the best fits.
From all best fits of all sets we derive the mean and the standard deviation of
the input parameters: \nh, \Av, FUV, and $\chi^2$ (Table\,\ref{modelsSources}).

\end{appendix}
\begin{appendix}

  \section{GMC properties}
  \label{AppendixA}

  For comparison we derived complementary properties, the star formation rate,
  the FUV field, the \XCO factor, the masses of the molecular and atomic
  material, and the optical extinction within the beam at the observed positions
  (Table~\ref{table:values}). Below, we explain in detail how these values have
  been derived. All errors are calculated using Gaussian error propagation. The
  particular uncertainties of individual measurements used to derive the
  complementary quantities are given below.

  \subsection{ Star formation rate (SFR) and total infrared luminosity
    (TIR)}
\label{TIRestimation}
To obtain an extinction free tracer of the SFR we combine the 24\,$\upmu$m
\citep{Tabatabaei2007} and H$\upalpha$ luminosity \citep{Hoopes2000} using
${\rm{SFR}} = 5.3\,10^{-42}[L({\rm{H}}\upalpha)+(0.031 \pm 0.006)
L(24\,\upmu{\rm{m}})]$ \citep{Calzetti2007b}. We assume errors of 7$\%$ for
24\,$\upmu$m (Spitzer Observers Manual v8.0) and 15$\%$ for H$\upalpha$.

Another measure of the SFR is the total infrared luminosity ($L_{\rm TIR}$)
emitted between 1$\,\mu$m and 1\,mm. We estimate it by adding the measured
brightnesses at 24, 70, 100, 160, 250, 350, 500\,$\mu$m, in W\,kpc$^{-2}$
weighted by the factors determined by \citet{Boquien2011}. We convert the
determined brightnesses to luminosities corresponding to the 28\,\arcsec\ beam
by multiplying with the beam size in kpc$^{-2}$. The errors on the derived TIR
are the $\sigma$ values of the fits given by \citet{Boquien2011}.

\subsection{Molecular line luminosities}
\label {lineLum}
We calculate the molecular line luminosities $L^{\prime}_{\rm HCN}$ and
$L^{\prime}_{\rm HCO^{+}}$ in units of K\,km\,s$^{-1}$\,pc$^2$ by multiplying
the integrated line intensities with the beam size in pc$^2$,
i.e. 1.5\,10$^{4}$\,pc$^{2}$. Errors follow from the measurement uncertainties
given in Table~\ref{table:values}

\subsection {FUV flux \Gn}

The FUV flux \Gn impinging on the dusty cloud surfaces can be estimated from
the emitted total infrared intensity assuming that all FUV photons are absorbed
by the dust. We use \mbox{\Gn\,$=\,4\pi\,0.5\,L_{\rm TIR}$} \citep[cf.\
e.g.][]{Mookerjea2011} with \Gn in units of the Habing field
\mbox{$1.6\,10^{-3}$ erg\,s$^{-1}$\,cm$^{2}$} \citep{Habing1968} and the TIR
intensity in units of erg\,s$^{-1}$\,cm$^{-2}$\,sr$^{-1}$. Errors follow
directly from the errors on $L_{\rm TIR}$. The FUV flux that we use to
compare with values deduced from the stacked spectrum is the average FUV flux
of all individual clouds, i.e. 21.6 $\pm$ 0.7\,\Gn.

\subsection{$X_{\rm CO}$ factor and atomic and molecular gas mass}
\label{XCODer}

To derive the masses of the molecular gas we first convert the
\mbox{CO\,1--0} integrated intensities into H$_2$ column densities ($N_{\rm
  H_2}$) via $N_{\rm H_2}~=~I_{\rm CO} X_{\rm CO}$. We assume that the
conversion factor $X_{\rm CO}$ (cf.~Table~\ref{table:values}) is a function of
total infrared luminosity as found by \citet{Israel1997} for a sample of
low-metallicity environments in the LMC and SMC and for two clouds in
M\,33 \citep[cf.][]{Leroy2011}.

\citet{Israel1997} derived $N_{\rm H_2}$ for CO clouds in the LMC,
the SMC, and other low-metallicity galaxies from their far-infrared surface
brightness and \HI\ column densities. He finds local $X_{\rm CO}$ values up to
a magnitude higher than the widely used Galactic value of 2\COU
\citep[e.g.][]{Dame2001}. In particular he studies the two bright \HII\ regions
NGC\,604 and NGC\,595 in M\,33, for which he derives X-factors of 22\,\COU and
12\,\COU. \citet{Israel1997} further finds a correlation between \XCO and the
TIR-flux. For NGC\,604 we find a TIR luminosity of 27.18 and 7.82
10$^6$ $L_{\sun}$ for NGC\,595 (cf.\,Sec.\,\ref{TIRestimation}). We estimate
the local \XCO for our seven clouds from their TIR by using the mean of the
\XCO/$L_{\rm TIR}$ ratio of NGC\,604 and NGC\,595 as a conversion factor, i.e.
\XCO = 1.17\,$10^{20}$ 10$^{-6}$ L$_{\sun}$ $L_{\rm TIR}$.  The derived
$X$-factors vary between 7 and 1.5 in units of \COU for the six positions with
HCN and \HCOp\ detections (Table\,\ref{table:values}).
\citet{Gratier2010} assume a constant X-factor of 4\,\COU.

The mass of the molecular gas $M_{\rm H_2}$ is estimated from the intensity of
the \mbox{\twCO\,1--0} line using the \XCO-factor derived in the previous
section and \mbox{$M_{\rm H_2} = 2\times1.36\,N_{\rm H_2}\,{\rm m}_{\rm p}\,A$}
with the proton mass m$_{\rm p}$, the beam size $A$. The factor two accounts for
the two protons of molecular hydrogen and 1.36 includes the contribution of
heavy elements to the total mass. Errors follow from the measurement
uncertainty of \twCO given in Table~\ref{table:values}.

The mass of the atomic gas is estimated from the intensity of the \HI line via
the column density \mbox{$N_{\HI} = 1.82\,\times\,10^{18}\,I_{\HI}$}
\citep{Rohlfs2000} assuming optically thin emission. The atomic gas mass
is calculated with the same formula as for the molecular mass without the
factor two accounting for the two protons in H$_2$. Errors
assume a 15\% calibration error on the HI data.

\subsection{Optical extinction \Av }
\label{optExt}

As shown by \citet[][]{Bohlin1978} a correlation between the amount of hydrogen
atom column density N(\HI + H$_2$) and the color excess exists; they estimate a
conversion factor of \mbox{$\langle$N(\HI +
  H$_2$)$\rangle$/E(B-V)\,$=\,5.8\,10^{21}$}
atoms\,cm$^{-2}$\,mag$^{-1}$. The color excess and the optical extinction \Av
are also linked. \cite{Bohlin1978} give a value for the interstellar average of
R~=~\Av/E(B-V)~=~3.1. Using the two conversion formula and the values for
N$_{\HI}$ and N$_{\rm H_2}$, derived as explained in the previous section, we
calculate an estimate of the optical extinction for the observed clouds per
beam. For clouds not filling the 115\,pc beam, local extinctions may be higher.
The errors given for \Av (cf.\,Table~\ref{table:values}) include the errors of
N(\HI) and N(H$_2$). We estimate the \Av for the stacked spectra by averaging
the values of the individual positions, i.e. 4.2~$\pm$~1.9\,mag.

\end{appendix}

\begin{appendix}
  \section{LTE column densities}
  \label{Appendix:LTE}

  \begin{table*}
    \caption{LTE column densities at the individual positions.}
    \label{Tab:Coldens}
    \centering
    \begin{tabular}{cccccccc}
      \hline
      \hline
					 & NO6      & GMC1     & GMC26    &	NO3      & GMC91	& NO1      & NO2          \\
      \hline
      $N$($^{12}$CO(1-0) [cm$^{-2}$])    & 1.19e+16	& 1.25e+16 & 1.14e+16 &	1.57e+16 & 3.53e+16	& 6.69e+15 & 2.27e+15     \\
      $N$($^{12}$CO(2-1) [cm$^{-2}$])    & 5.82e+15	& 6.89e+15 & 4.60e+15 &	6.16e+15 & 1.29e+16	& 4.08e+15 & 5.05e+14     \\
      $N$($^{13}$CO(1-0) [cm$^{-2}$])    & 1.67e+15	& 2.96e+15 & 1.90e+15 &	2.74e+15 & 5.87e+15	& 1.31e+15 & 4.54e+14     \\
      $N$(HCO$^+$ [cm$^{-2}$])           & 4.13e+11	& 3.83e+11 & 1.32e+11 &	2.38e+11 & 1.89e+11	& 1.54e+11 & $<$ 2.29e+10 \\
      $N$(HCN [cm$^{-2}$])               & 8.64e+11	& 1.79e+12 & 5.72e+11 &	6.38e+11 & 6.78e+11	& 5.88e+11 & 2.59e+11     \\
      \hline
      $N$(H$_2$ [cm$^{-2}$]) - LTE       & 1.18e+21	& 2.09e+21 & 1.34e+21 &	1.94e+21 & 4.14e+21	& 9.28e+20 & 3.21e+20     \\
      $N$(H$_2$ [cm$^{-1}$]) - X-factor  & 2.98e+21	& 3.95e+21 & 9.32e+20 &	2.45e+21 & 2.55e+21	& 6.51e+20 & 4.00e+19     \\
      \hline
      log $N$($^{12}$CO(1-0))/$N$(H$_2$) & -5.00	& -5.22    & -5.07    &	-5.09    & -5.07	& -5.14    & -5.15        \\
      log $N$($^{12}$CO(2-1))/$N$(H$_2$) & -5.31	& -5.48    & -5.46    &	-5.50    & -5.51	& -5.36    & -5.80        \\
      log $N$($^{13}$CO(1-0))/$N$(H$_2$) & -5.85	& -5.85    & -5.85    &	-5.85    & -5.85	& -5.85    & -5.85        \\
      log $N$(HCO$^+$)/$N$(H$_2$)        & -9.46	& -9.74    & -10.01   &	-9.91    & -10.34	& -9.78    & $<$ -10.15   \\
      log $N$(HCN)/$N$(H$_2$)            & -9.14	& -9.07    & -9.37    &	-9.48    & -9.79	& -9.20    & -9.09        \\
      $N$(HCO$^+$)/$N$(HCN)              & 0.48     & 0.21     & 0.23     &	0.37     & 0.28     & 0.26     & $<$ 0.09     \\
      T$_{\rm ex}$ [K]                   & 23       & 25       & 22       &	22       & 20       & 23       & 19           \\
      \hline\end{tabular}
  \end{table*}

  We estimate column densities from the integrated intensities, assuming LTE
  and optically thin emission:
  \begin{equation}
    \label{formula:coldens}
    N = \frac{3h}{8 \pi^3 \mu^2} \frac{Z}{J} \frac{exp(\frac{h \nu}{k
	T_{ex}})}{[1 - exp(-\frac{h \nu}{k T_{ex}})]}
    (\mathcal{J}_{\nu}(T_{ex}) -
    \mathcal{J}_{\nu}(T_{BG}))^{-1} \int{T_{mb} d \nu} ,
  \end{equation}

  with:
  \begin{equation}
    \label{formula:J}
    \mathcal{J}_{\nu}(T) = \frac{h\nu}{k} \frac{1}{e^{h\nu/kT_{ex}}-1} ,
  \end{equation}
  where $Z$ the partition function, $J$ the rotational quantum number of the
  upper level, $\mu$ the dipole moment, $T_{\rm ex}$ the excitation
  temperature, $T_{\rm BG}$ the temperature of the cosmic background, and
  $T_{\rm mb}$ the main beam temperature of the specific
  line. Table~\ref{Tab:moleculeParameter} lists frequency, $\mu$, \mbox{$E_{\rm
      u}/{\rm k}_b$}, and the critical densities of the observed
  transitions. The partition function is taken from the ``Cologne Database for
  Molecular Spectroscopy'' (CDMS) \citep{Muller2001,Muller2005}. The value of
  $Z$ depends on $T_{\rm ex}$. The CDMS lists values for discrete temperatures
  in the range from 2.75 to 500\,K for every molecule. We use these values to
  interpolate $Z$ for the excitation temperature we adopt at a specific
  position.

  \begin{table}
%
    % the value of the Boltzmann constant k expressed in wave-number units per
    % kelvin is 0.695 035 6(12) cm-1/K from
    % http://physics.nist.gov/Pubs/AtSpec/node01.html
%
    \caption{Molecule parameters }
    \label{Tab:moleculeParameter}
    \centering
    \begin{tabular}{ccccc}
      \hline \hline
      & $\nu$ & $\mu$ & $E_{\rm u}/{\rm k}_{\rm b}$   & $n_{\rm cr}$\tablefootmark{a} \\
      & [GHz]	& [Debye]   & [K] & [cm$^{-3}$] \\
      \hline
      C$_2$H(1--0, 3/2-1/2) & 87.2841050  & 0.77 & 4.19 & 1 \, 10$^5$\\
      HCN\,1--0             & 88.6316022  & 2.99 & 4.25 & 2 \, 10$^5$\\
      \HCOp\,1--0           & 89.1885247  & 3.90 & 4.28 & 3 \, 10$^4$\\
      \HOCp\,1--0           & 89.4874140  & 2.77 & 4.29 & 3 \, 10$^4$\\
      HNC\,1--0             & 90.6635680  & 3.05 & 4.35 & 2 \, 10$^5$\\
      \thCO\,1--0           & 110.2013543 & 0.11 & 5.28 & 4 \, 10$^2$\\
      \twCO\,1--0           & 115.2712018 & 0.11 & 5.53 & 4 \, 10$^2$\\
      \hline
    \end{tabular}
    \tablefoot{\tablefoottext{a}{for collisions with H$_2$ neglecting opacity effects.}}
  \end{table}

  To estimate the excitation temperature, we derive the dust temperature,
  assuming that $T_{\rm ex}$, the kinetic temperature $T_{\rm kin}$, and the
  dust temperature $T_{\rm dust}$ are coupled and equal. To estimate $T_{\rm
    dust}$ we obtain the dust SEDs for all our pointings from the Herschel and
  Spitzer observations at 500, 350, 250, 160, 100, and 24 \,$\mu$m 28\arcsec
  resolution and fit two-component graybody models to the data. Each component
  is described by \mbox{$ S_{\nu} = B(\nu, T ) \tau_{\nu} = B(\nu, T )
    \kappa_{\nu} M_d / D^2$} , assuming optically thin emission, with the flux
  S$_{\nu}$ , the Planck function B$_{\nu}$ , the opacity $\tau_{\nu}$ , the
  dust mass $M_d$ , the distance D, and the dust absorption coefficient
  \mbox{$\kappa_{\nu}$ = 0.4($\nu$/(250 GHz))$^{\beta}$ cm$^2$ g$^{-1}$}
  \citep{Kruegel1994,Kruegel2003}, $\beta$ is the dust emissivity index assumed
  to be 1.5.

  T$_{\rm dust}$ is found to vary only slightly, between 20\,K in the
  northernmost position observed, GMC no2, and 26\,K in the nuclear GMC, GMC1
  (cf.\,Table~\ref{Tab:Coldens}). We assume that $T_{\rm ex}$ equals the
  temperature of the cold dust component.

  The \mbox{CCH\,1--0} transition is split into six hyperfine components. In
  the stacked spectrum, we detected the \mbox{3/2--1/2}, \mbox{2--1}
  transition, which is the strongest one. To determine LTE column densities of
  CCH, we estimate the total integrated intensity of the \mbox{CCH\,1--0} line
  including all hyperfine components by dividing the result of a Gaussian fit
  to the detected component by its relative strength, i.e. 0.416
  \citep{Padovani2009}.

  The column density of H$_2$ is estimated from \mbox{\thCO\,1--0} , assuming a
  \twCO/\thCO\ abundance ratio of 60 as typically found in the Milky Way
  \citep{Langer1993} and a H$_2$/CO abundance ratio of $8.5\,10^{-5}$, also
  found for the Milky Way \citep[][]{Frerking1982}. From this we derive the
  relative abundances of the different species with respect to $N_{\rm
    H_2}$. The derived column densities, as well as the relative abundances, are
  given in Table~\ref{Tab:Coldens} for individual clouds and in
  Table~\ref{table:stackedTable} for the stacked spectra. For comparison we
  give in Table~\ref{Tab:Coldens} the values of $N_{\rm H_2}$ that we derive
  from the \mbox{\twCO\,1--0} lines using the individual $X_{\rm CO}$-factors
  derived for every cloud (see above). After comparing both methods to obtain the
  column density of H$_2$, we find that the results are consistent within a
  factor of 2.5 excluding position no2. The values of $N_{\rm H_2}$ for the
  stacked values in Table~\ref{table:stackedTable} are deduced using an average
  over the individual \XCO factor of the observed GMCs of 3.0 \COU.

\end{appendix}

\begin{appendix}
\section{PDR Model results}

The modeled abundances and intensity ratios of the ground-state transition of
HCN, HCO, and \twCO using the Meudon PDR code for subsolar and solar
metallicities are shown in Tables \ref{table:PDRResultsSubSolar} and
\ref{table:PDRResultsSolar}. Both tables are subdivided in three blocks one for
each modeled radiation field of \mbox{\Gn\,$=\,10, 50, {\rm and}\ 100$} in
Habing units, which illuminates the modeled clouds from both sides. The other two
input parameters that have been varied are the density with values of
 \nh~=~0.1, 0.5, 1, 5, 10, 50, 10$^2$ $\times$
10$^4$\,cm$^{-3}$ and the optical extinctions with values of \mbox{\Av~=~2, 4, 6,
  10, 16, 26, 40, and 50\,mag}.

\longtab{1}{
\begin{longtable}{@{}rr|rrrrrr}
\caption{\label{table:PDRResultsSubSolar} Results of the PDR models with subsolar metallicities for M\,33
  (cf. Sect.~\ref{Sec:PDR}).}\\
\hline
\hline
\Av & \nh  & $X$(\HCOp)\tablefootmark{a} & $X$(HCN)\tablefootmark{a} & $X$(CO)\tablefootmark{a} & HCN/CO& \HCOp/CO & \HCOp/HCN\\
(mag.) & (cm$^{-3}$)   & & & & \% & \% &\\
\hline
\endfirsthead
\caption{Continued} \\
\hline
\Av & \nh  & $X$(\HCOp)\tablefootmark{a} & $X$(HCN)\tablefootmark{a} & $X$(CO)\tablefootmark{a} & HCN/CO& \HCOp/CO & \HCOp/HCN\\
(mag.) & (cm$^{-3}$)   & & & & \% & \% & \\
\hline
\hline
\endhead
\hline
\endfoot
\hline
\endlastfoot
\hline
\multicolumn{8}{c}{ \Gn = 10}\\
\hline
2.0 & 1\,10$^{3}$ & -10.8 & -11.7 & -7.4 & 1.11 & 19.57 & 17.63 \\
2.0 & 5\,10$^{3}$ & -11.2 & -10.7 & -6.1 & 0.65 & 0.47 & 0.73 \\
2.0 & 1\,10$^{4}$ & -11.3 & -10.4 & -5.6 & 0.40 & 0.14 & 0.34 \\
2.0 & 5\,10$^{4}$ & -11.1 & -9.9 & -4.6 & 0.22 & 0.04 & 0.19 \\
2.0 & 1\,10$^{5}$ & -11.0 & -9.7 & -4.4 & 0.26 & 0.04 & 0.14 \\
2.0 & 5\,10$^{5}$ & -11.2 & -9.3 & -4.1 & 0.42 & 0.02 & 0.04 \\
2.0 & 1\,10$^{6}$ & -11.2 & -9.2 & -4.1 & 0.47 & 0.01 & 0.02 \\
4.0 & 1\,10$^{3}$ & -10.8 & -10.9 & -5.7 & 0.19 & 0.46 & 2.43 \\
4.0 & 5\,10$^{3}$ & -11.0 & -10.3 & -4.7 & 0.13 & 0.06 & 0.43 \\
4.0 & 1\,10$^{4}$ & -11.0 & -10.1 & -4.4 & 0.18 & 0.05 & 0.27 \\
4.0 & 5\,10$^{4}$ & -11.1 & -9.7 & -4.2 & 0.37 & 0.04 & 0.11 \\
4.0 & 1\,10$^{5}$ & -11.1 & -9.5 & -4.1 & 0.48 & 0.04 & 0.08 \\
4.0 & 5\,10$^{5}$ & -11.3 & -9.3 & -4.0 & 0.73 & 0.02 & 0.03 \\
4.0 & 1\,10$^{6}$ & -11.4 & -9.3 & -4.0 & 0.79 & 0.01 & 0.02 \\
6.0 & 1\,10$^{3}$ & -10.5 & -10.5 & -5.4 & 0.39 & 0.68 & 1.73 \\
6.0 & 5\,10$^{3}$ & -10.7 & -10.0 & -4.5 & 0.30 & 0.16 & 0.54 \\
6.0 & 1\,10$^{4}$ & -10.6 & -9.9 & -4.3 & 0.38 & 0.15 & 0.40 \\
6.0 & 5\,10$^{4}$ & -10.7 & -9.6 & -4.1 & 0.61 & 0.12 & 0.20 \\
6.0 & 1\,10$^{5}$ & -10.8 & -9.5 & -4.1 & 0.71 & 0.10 & 0.14 \\
6.0 & 5\,10$^{5}$ & -11.1 & -9.4 & -4.0 & 0.89 & 0.05 & 0.06 \\
6.0 & 1\,10$^{6}$ & -11.2 & -9.4 & -4.0 & 0.91 & 0.04 & 0.04 \\
10.0 & 1\,10$^{3}$ & -9.2 & -9.0 & -4.5 & 6.10 & 8.22 & 1.35 \\
10.0 & 5\,10$^{3}$ & -9.4 & -9.2 & -4.2 & 2.36 & 3.14 & 1.33 \\
10.0 & 1\,10$^{4}$ & -9.6 & -9.3 & -4.1 & 1.79 & 2.02 & 1.13 \\
10.0 & 5\,10$^{4}$ & -10.1 & -9.4 & -4.0 & 1.62 & 0.80 & 0.49 \\
10.0 & 1\,10$^{5}$ & -10.3 & -9.3 & -4.0 & 1.61 & 0.56 & 0.35 \\
10.0 & 5\,10$^{5}$ & -10.6 & -9.4 & -4.0 & 1.33 & 0.24 & 0.18 \\
10.0 & 1\,10$^{6}$ & -10.8 & -9.5 & -4.0 & 1.21 & 0.17 & 0.14 \\
16.0 & 1\,10$^{3}$ & -8.6 & -8.3 & -4.2 & 14.29 & 15.42 & 1.08 \\
16.0 & 5\,10$^{3}$ & -9.1 & -8.6 & -4.1 & 7.98 & 7.25 & 0.91 \\
16.0 & 1\,10$^{4}$ & -9.3 & -8.6 & -4.0 & 7.42 & 5.05 & 0.68 \\
16.0 & 5\,10$^{4}$ & -9.8 & -8.7 & -4.0 & 6.71 & 2.20 & 0.33 \\
16.0 & 1\,10$^{5}$ & -10.0 & -8.9 & -4.0 & 5.86 & 1.60 & 0.27 \\
16.0 & 5\,10$^{5}$ & -10.3 & -9.2 & -4.0 & 3.30 & 0.74 & 0.22 \\
16.0 & 1\,10$^{6}$ & -10.5 & -9.3 & -3.9 & 2.52 & 0.52 & 0.20 \\
26.0 & 1\,10$^{3}$ & -8.4 & -8.1 & -4.1 & 15.30 & 17.49 & 1.14 \\
26.0 & 5\,10$^{3}$ & -8.9 & -8.4 & -4.0 & 10.34 & 10.55 & 1.02 \\
26.0 & 1\,10$^{4}$ & -9.2 & -8.4 & -4.0 & 10.14 & 8.30 & 0.82 \\
26.0 & 5\,10$^{4}$ & -9.7 & -8.5 & -4.0 & 10.81 & 4.32 & 0.40 \\
26.0 & 1\,10$^{5}$ & -9.9 & -8.7 & -4.0 & 10.45 & 3.24 & 0.31 \\
26.0 & 5\,10$^{5}$ & -10.2 & -9.1 & -3.9 & 6.79 & 1.56 & 0.23 \\
26.0 & 1\,10$^{6}$ & -10.4 & -9.2 & -3.9 & 5.06 & 1.10 & 0.22 \\
40.0 & 1\,10$^{3}$ & -8.3 & -8.1 & -4.0 & 15.52 & 19.00 & 1.23 \\
40.0 & 5\,10$^{3}$ & -8.9 & -8.4 & -4.0 & 11.44 & 12.98 & 1.14 \\
40.0 & 1\,10$^{4}$ & -9.1 & -8.4 & -4.0 & 11.59 & 11.11 & 0.96 \\
40.0 & 5\,10$^{4}$ & -9.6 & -8.5 & -3.9 & 13.63 & 6.79 & 0.50 \\
40.0 & 1\,10$^{5}$ & -9.8 & -8.6 & -3.9 & 14.05 & 5.25 & 0.37 \\
40.0 & 5\,10$^{5}$ & -10.2 & -9.0 & -3.9 & 10.59 & 2.63 & 0.25 \\
40.0 & 1\,10$^{6}$ & -10.3 & -9.2 & -3.9 & 8.08 & 1.86 & 0.23 \\
50.0 & 1\,10$^{3}$ & -8.2 & -8.1 & -4.0 & 15.51 & 19.81 & 1.28 \\
50.0 & 5\,10$^{3}$ & -8.8 & -8.3 & -4.0 & 11.89 & 14.17 & 1.19 \\
50.0 & 1\,10$^{4}$ & -9.1 & -8.3 & -4.0 & 12.23 & 12.53 & 1.02 \\
50.0 & 5\,10$^{4}$ & -9.6 & -8.4 & -3.9 & 14.97 & 8.28 & 0.55 \\
50.0 & 1\,10$^{5}$ & -9.8 & -8.6 & -3.9 & 15.82 & 6.54 & 0.41 \\
50.0 & 5\,10$^{5}$ & -10.1 & -9.0 & -3.9 & 12.80 & 3.35 & 0.26 \\
50.0 & 1\,10$^{6}$ & -10.3 & -9.2 & -3.9 & 9.96 & 2.39 & 0.24 \\
\hline
\multicolumn{8}{c}{ \Gn = 50}\\
\hline
2.0 & 1\,10$^{3}$ & -11.1 & -13.0 & -8.7 & 1.19 & 219.55 & 185.29 \\
2.0 & 5\,10$^{3}$ & -11.3 & -11.7 & -7.6 & 2.38 & 15.95 & 6.71 \\
2.0 & 1\,10$^{4}$ & -11.4 & -11.2 & -7.1 & 2.29 & 4.48 & 1.95 \\
2.0 & 5\,10$^{4}$ & -11.4 & -10.4 & -5.8 & 0.98 & 0.37 & 0.38 \\
2.0 & 1\,10$^{5}$ & -10.9 & -10.2 & -5.2 & 0.53 & 0.33 & 0.61 \\
2.0 & 5\,10$^{5}$ & -11.0 & -9.6 & -4.5 & 0.27 & 0.04 & 0.13 \\
2.0 & 1\,10$^{6}$ & -11.1 & -9.4 & -4.3 & 0.29 & 0.02 & 0.06 \\
4.0 & 1\,10$^{3}$ & -11.0 & -11.6 & -6.7 & 0.24 & 2.34 & 9.64 \\
4.0 & 5\,10$^{3}$ & -11.2 & -10.8 & -5.3 & 0.11 & 0.09 & 0.75 \\
4.0 & 1\,10$^{4}$ & -11.3 & -10.5 & -4.9 & 0.10 & 0.04 & 0.40 \\
4.0 & 5\,10$^{4}$ & -11.2 & -10.0 & -4.4 & 0.18 & 0.03 & 0.16 \\
4.0 & 1\,10$^{5}$ & -11.0 & -9.8 & -4.3 & 0.24 & 0.04 & 0.18 \\
4.0 & 5\,10$^{5}$ & -11.2 & -9.5 & -4.1 & 0.40 & 0.02 & 0.05 \\
4.0 & 1\,10$^{6}$ & -11.3 & -9.4 & -4.1 & 0.45 & 0.01 & 0.03 \\
6.0 & 1\,10$^{3}$ & -10.8 & -11.0 & -5.8 & 0.20 & 0.57 & 2.81 \\
6.0 & 5\,10$^{3}$ & -11.0 & -10.4 & -4.8 & 0.14 & 0.09 & 0.59 \\
6.0 & 1\,10$^{4}$ & -11.0 & -10.2 & -4.5 & 0.18 & 0.07 & 0.38 \\
6.0 & 5\,10$^{4}$ & -11.1 & -9.8 & -4.2 & 0.34 & 0.05 & 0.16 \\
6.0 & 1\,10$^{5}$ & -11.0 & -9.7 & -4.1 & 0.41 & 0.06 & 0.15 \\
6.0 & 5\,10$^{5}$ & -11.2 & -9.5 & -4.1 & 0.57 & 0.03 & 0.05 \\
6.0 & 1\,10$^{6}$ & -11.3 & -9.4 & -4.0 & 0.59 & 0.02 & 0.03 \\
10.0 & 1\,10$^{3}$ & -10.2 & -10.0 & -5.1 & 1.06 & 1.41 & 1.33 \\
10.0 & 5\,10$^{3}$ & -9.8 & -9.6 & -4.3 & 1.02 & 1.54 & 1.51 \\
10.0 & 1\,10$^{4}$ & -9.9 & -9.7 & -4.2 & 0.84 & 1.08 & 1.29 \\
10.0 & 5\,10$^{4}$ & -10.3 & -9.7 & -4.1 & 0.74 & 0.48 & 0.65 \\
10.0 & 1\,10$^{5}$ & -10.4 & -9.6 & -4.0 & 0.75 & 0.36 & 0.48 \\
10.0 & 5\,10$^{5}$ & -10.8 & -9.6 & -4.0 & 0.77 & 0.14 & 0.18 \\
10.0 & 1\,10$^{6}$ & -10.9 & -9.5 & -4.0 & 0.74 & 0.09 & 0.12 \\
16.0 & 1\,10$^{3}$ & -8.7 & -8.4 & -4.3 & 13.57 & 14.65 & 1.08 \\
16.0 & 5\,10$^{3}$ & -9.2 & -8.8 & -4.1 & 6.49 & 6.21 & 0.96 \\
16.0 & 1\,10$^{4}$ & -9.4 & -8.8 & -4.1 & 5.54 & 4.16 & 0.75 \\
16.0 & 5\,10$^{4}$ & -9.9 & -9.0 & -4.0 & 4.23 & 1.80 & 0.43 \\
16.0 & 1\,10$^{5}$ & -10.0 & -9.1 & -4.0 & 3.43 & 1.31 & 0.38 \\
16.0 & 5\,10$^{5}$ & -10.4 & -9.4 & -4.0 & 1.79 & 0.55 & 0.31 \\
16.0 & 1\,10$^{6}$ & -10.5 & -9.5 & -4.0 & 1.38 & 0.36 & 0.26 \\
26.0 & 1\,10$^{3}$ & -8.4 & -8.2 & -4.1 & 15.23 & 17.30 & 1.14 \\
26.0 & 5\,10$^{3}$ & -9.0 & -8.5 & -4.0 & 9.56 & 9.90 & 1.04 \\
26.0 & 1\,10$^{4}$ & -9.2 & -8.5 & -4.0 & 8.88 & 7.56 & 0.85 \\
26.0 & 5\,10$^{4}$ & -9.7 & -8.7 & -4.0 & 8.42 & 3.97 & 0.47 \\
26.0 & 1\,10$^{5}$ & -9.8 & -8.8 & -4.0 & 7.58 & 2.97 & 0.39 \\
26.0 & 5\,10$^{5}$ & -10.2 & -9.2 & -4.0 & 4.22 & 1.30 & 0.31 \\
26.0 & 1\,10$^{6}$ & -10.3 & -9.3 & -4.0 & 3.02 & 0.85 & 0.28 \\
40.0 & 1\,10$^{3}$ & -8.3 & -8.1 & -4.0 & 15.47 & 18.89 & 1.22 \\
40.0 & 5\,10$^{3}$ & -8.9 & -8.4 & -4.0 & 10.74 & 12.44 & 1.16 \\
40.0 & 1\,10$^{4}$ & -9.1 & -8.4 & -4.0 & 10.36 & 10.40 & 1.00 \\
40.0 & 5\,10$^{4}$ & -9.6 & -8.6 & -4.0 & 11.18 & 6.47 & 0.58 \\
40.0 & 1\,10$^{5}$ & -9.8 & -8.7 & -4.0 & 10.91 & 5.01 & 0.46 \\
40.0 & 5\,10$^{5}$ & -10.1 & -9.1 & -3.9 & 6.97 & 2.27 & 0.33 \\
40.0 & 1\,10$^{6}$ & -10.3 & -9.3 & -3.9 & 5.03 & 1.50 & 0.30 \\
50.0 & 1\,10$^{3}$ & -8.2 & -8.1 & -4.0 & 15.46 & 19.72 & 1.28 \\
50.0 & 5\,10$^{3}$ & -8.8 & -8.4 & -4.0 & 11.22 & 13.68 & 1.22 \\
50.0 & 1\,10$^{4}$ & -9.1 & -8.4 & -4.0 & 10.99 & 11.84 & 1.08 \\
50.0 & 5\,10$^{4}$ & -9.6 & -8.5 & -4.0 & 12.46 & 8.00 & 0.64 \\
50.0 & 1\,10$^{5}$ & -9.7 & -8.7 & -4.0 & 12.57 & 6.30 & 0.50 \\
50.0 & 5\,10$^{5}$ & -10.1 & -9.1 & -3.9 & 8.65 & 2.92 & 0.34 \\
50.0 & 1\,10$^{6}$ & -10.2 & -9.2 & -3.9 & 6.33 & 1.95 & 0.31 \\
\hline
\multicolumn{8}{c}{ \Gn = 100}\\
\hline
2.0 & 1\,10$^{3}$ & -11.2 & -13.6 & -9.3 & 1.08 & 536.12 & 498.10 \\
2.0 & 5\,10$^{3}$ & -11.4 & -12.4 & -8.4 & 2.97 & 69.85 & 23.49 \\
2.0 & 1\,10$^{4}$ & -11.4 & -11.9 & -7.8 & 3.43 & 23.10 & 6.74 \\
2.0 & 5\,10$^{4}$ & -11.5 & -10.9 & -6.5 & 2.10 & 1.76 & 0.84 \\
2.0 & 1\,10$^{5}$ & -11.0 & -10.6 & -6.0 & 1.21 & 1.66 & 1.37 \\
2.0 & 5\,10$^{5}$ & -10.7 & -10.0 & -4.9 & 0.28 & 0.16 & 0.56 \\
2.0 & 1\,10$^{6}$ & -11.0 & -9.8 & -4.6 & 0.22 & 0.03 & 0.13 \\
4.0 & 1\,10$^{3}$ & -11.1 & -12.1 & -7.2 & 0.29 & 6.97 & 23.75 \\
4.0 & 5\,10$^{3}$ & -11.3 & -11.1 & -5.9 & 0.18 & 0.24 & 1.31 \\
4.0 & 1\,10$^{4}$ & -11.4 & -10.7 & -5.3 & 0.13 & 0.07 & 0.58 \\
4.0 & 5\,10$^{4}$ & -11.3 & -10.2 & -4.6 & 0.14 & 0.03 & 0.20 \\
4.0 & 1\,10$^{5}$ & -11.1 & -10.0 & -4.4 & 0.18 & 0.04 & 0.23 \\
4.0 & 5\,10$^{5}$ & -10.9 & -9.6 & -4.2 & 0.28 & 0.03 & 0.11 \\
4.0 & 1\,10$^{6}$ & -11.2 & -9.5 & -4.2 & 0.32 & 0.01 & 0.04 \\
6.0 & 1\,10$^{3}$ & -11.0 & -11.2 & -6.0 & 0.15 & 0.57 & 3.73 \\
6.0 & 5\,10$^{3}$ & -11.1 & -10.6 & -4.9 & 0.10 & 0.07 & 0.68 \\
6.0 & 1\,10$^{4}$ & -11.1 & -10.3 & -4.6 & 0.13 & 0.05 & 0.43 \\
6.0 & 5\,10$^{4}$ & -11.2 & -10.0 & -4.3 & 0.24 & 0.04 & 0.17 \\
6.0 & 1\,10$^{5}$ & -11.1 & -9.8 & -4.2 & 0.29 & 0.05 & 0.17 \\
6.0 & 5\,10$^{5}$ & -11.0 & -9.6 & -4.1 & 0.41 & 0.04 & 0.09 \\
6.0 & 1\,10$^{6}$ & -11.3 & -9.5 & -4.1 & 0.44 & 0.02 & 0.04 \\
10.0 & 1\,10$^{3}$ & -10.3 & -10.1 & -5.3 & 0.92 & 1.27 & 1.39 \\
10.0 & 5\,10$^{3}$ & -10.0 & -9.8 & -4.4 & 0.72 & 0.93 & 1.30 \\
10.0 & 1\,10$^{4}$ & -10.1 & -9.8 & -4.3 & 0.64 & 0.74 & 1.16 \\
10.0 & 5\,10$^{4}$ & -10.4 & -9.7 & -4.1 & 0.59 & 0.37 & 0.63 \\
10.0 & 5\,10$^{5}$ & -10.8 & -9.6 & -4.0 & 0.58 & 0.12 & 0.21 \\
10.0 & 1\,10$^{6}$ & -11.0 & -9.6 & -4.0 & 0.56 & 0.07 & 0.12 \\
16.0 & 1\,10$^{3}$ & -8.8 & -8.5 & -4.3 & 13.03 & 14.15 & 1.08 \\
16.0 & 5\,10$^{3}$ & -9.2 & -8.9 & -4.2 & 5.81 & 5.81 & 1.00 \\
16.0 & 1\,10$^{4}$ & -9.4 & -8.9 & -4.1 & 4.71 & 3.81 & 0.81 \\
16.0 & 5\,10$^{4}$ & -9.9 & -9.1 & -4.0 & 3.32 & 1.64 & 0.50 \\
16.0 & 1\,10$^{5}$ & -10.0 & -9.2 & -4.0 & 2.62 & 1.19 & 0.45 \\
16.0 & 5\,10$^{5}$ & -10.4 & -9.5 & -4.0 & 1.35 & 0.49 & 0.36 \\
16.0 & 1\,10$^{6}$ & -10.5 & -9.5 & -4.0 & 1.04 & 0.30 & 0.29 \\
26.0 & 1\,10$^{3}$ & -8.4 & -8.2 & -4.1 & 15.16 & 17.18 & 1.13 \\
26.0 & 5\,10$^{3}$ & -9.0 & -8.5 & -4.1 & 9.34 & 9.75 & 1.04 \\
26.0 & 1\,10$^{4}$ & -9.2 & -8.5 & -4.0 & 8.40 & 7.31 & 0.87 \\
26.0 & 5\,10$^{4}$ & -9.7 & -8.7 & -4.0 & 7.50 & 3.84 & 0.51 \\
26.0 & 1\,10$^{5}$ & -9.8 & -8.9 & -4.0 & 6.53 & 2.87 & 0.44 \\
26.0 & 5\,10$^{5}$ & -10.2 & -9.2 & -4.0 & 3.42 & 1.21 & 0.35 \\
26.0 & 1\,10$^{6}$ & -10.3 & -9.4 & -4.0 & 2.42 & 0.77 & 0.32 \\
40.0 & 1\,10$^{3}$ & -8.3 & -8.1 & -4.1 & 15.43 & 18.82 & 1.22 \\
40.0 & 5\,10$^{3}$ & -8.9 & -8.4 & -4.0 & 10.58 & 12.39 & 1.17 \\
40.0 & 1\,10$^{4}$ & -9.1 & -8.4 & -4.0 & 9.92 & 10.21 & 1.03 \\
40.0 & 5\,10$^{4}$ & -9.6 & -8.6 & -4.0 & 10.22 & 6.38 & 0.62 \\
40.0 & 1\,10$^{5}$ & -9.7 & -8.8 & -4.0 & 9.69 & 4.92 & 0.51 \\
40.0 & 5\,10$^{5}$ & -10.1 & -9.1 & -4.0 & 5.80 & 2.14 & 0.37 \\
40.0 & 1\,10$^{6}$ & -10.2 & -9.3 & -3.9 & 4.12 & 1.37 & 0.33 \\
50.0 & 1\,10$^{3}$ & -8.3 & -8.1 & -4.0 & 15.49 & 19.71 & 1.27 \\
50.0 & 5\,10$^{3}$ & -8.8 & -8.4 & -4.0 & 11.06 & 13.65 & 1.23 \\
50.0 & 1\,10$^{4}$ & -9.1 & -8.4 & -4.0 & 10.55 & 11.67 & 1.11 \\
50.0 & 5\,10$^{4}$ & -9.5 & -8.6 & -4.0 & 11.47 & 7.93 & 0.69 \\
50.0 & 1\,10$^{5}$ & -9.7 & -8.7 & -4.0 & 11.28 & 6.23 & 0.55 \\
50.0 & 5\,10$^{5}$ & -10.1 & -9.1 & -3.9 & 7.26 & 2.76 & 0.38 \\
50.0 & 1\,10$^{6}$ & -10.2 & -9.3 & -3.9 & 5.23 & 1.78 & 0.34 \\

\end{longtable}

\tablefoot{\tablefoottext{a}{logarithmic relative abundance
    $X$(mol)$=\log(N(\rm mol)/$$N$$({\rm H}_2))$}} }

\longtab{2}{
\begin{longtable}{@{}rr|rrrrrr}
\caption{\label{table:PDRResultsSolar} Results of the PDR models with solar metallicities.}\\
\hline
\hline
 \Av  & \nh  & $X$(\HCOp)\tablefootmark{a} & X(HCN)\tablefootmark{a} & X(CO)\tablefootmark{a} & HCN/CO& \HCOp/CO & \HCOp/HCN\\
    (mag.) & (cm$^{-3}$)   & & & & \% & \% &\\
\hline
\endfirsthead
\caption{Continued} \\
\hline
\Av & \nh  & $X$(\HCOp)\tablefootmark{a} & X(HCN)\tablefootmark{a} & X(CO)\tablefootmark{a} & HCN/CO& \HCOp/CO & \HCOp/HCN\\
(mag.) & (cm$^{-3}$)   & & & & \% & \% & \\
\hline
\hline
\endhead
\hline
\endfoot
\hline
\endlastfoot
\hline
\multicolumn{8}{c}{ \Gn = 10}\\
\hline
2.0 & 1\,10$^{3}$ & -11.6 & -11.0 & -6.5 & 0.84 & 0.45 & 0.54 \\
2.0 & 5\,10$^{3}$ & -11.6 & -10.4 & -5.1 & 0.15 & 0.02 & 0.12 \\
2.0 & 1\,10$^{4}$ & -11.5 & -10.2 & -4.6 & 0.11 & 0.01 & 0.10 \\
2.0 & 5\,10$^{4}$ & -11.3 & -9.6 & -3.8 & 0.18 & 0.01 & 0.07 \\
2.0 & 1\,10$^{5}$ & -11.1 & -9.3 & -3.7 & 0.29 & 0.02 & 0.06 \\
2.0 & 5\,10$^{5}$ & -11.2 & -8.9 & -3.5 & 0.59 & 0.01 & 0.02 \\
2.0 & 1\,10$^{6}$ & -11.2 & -8.8 & -3.4 & 0.68 & 0.01 & 0.01 \\
4.0 & 1\,10$^{3}$ & -11.4 & -10.5 & -4.7 & 0.12 & 0.03 & 0.29 \\
4.0 & 5\,10$^{3}$ & -11.1 & -9.9 & -3.8 & 0.19 & 0.03 & 0.16 \\
4.0 & 1\,10$^{4}$ & -11.0 & -9.7 & -3.7 & 0.29 & 0.03 & 0.12 \\
4.0 & 5\,10$^{4}$ & -11.1 & -9.3 & -3.5 & 0.64 & 0.03 & 0.05 \\
4.0 & 1\,10$^{5}$ & -11.1 & -9.1 & -3.5 & 0.83 & 0.03 & 0.04 \\
4.0 & 5\,10$^{5}$ & -11.3 & -8.9 & -3.4 & 1.22 & 0.02 & 0.01 \\
4.0 & 1\,10$^{6}$ & -11.3 & -8.8 & -3.4 & 1.33 & 0.01 & 0.01 \\
6.0 & 1\,10$^{3}$ & -10.8 & -10.1 & -4.4 & 0.30 & 0.15 & 0.51 \\
6.0 & 5\,10$^{3}$ & -10.5 & -9.7 & -3.7 & 0.48 & 0.16 & 0.34 \\
6.0 & 1\,10$^{4}$ & -10.5 & -9.5 & -3.6 & 0.65 & 0.15 & 0.23 \\
6.0 & 5\,10$^{4}$ & -10.7 & -9.2 & -3.5 & 1.03 & 0.10 & 0.10 \\
6.0 & 1\,10$^{5}$ & -10.7 & -9.1 & -3.4 & 1.19 & 0.09 & 0.08 \\
6.0 & 5\,10$^{5}$ & -11.0 & -9.0 & -3.4 & 1.46 & 0.04 & 0.03 \\
6.0 & 1\,10$^{6}$ & -11.1 & -9.0 & -3.4 & 1.52 & 0.03 & 0.02 \\
10.0 & 1\,10$^{3}$ & -9.4 & -9.2 & -3.7 & 2.81 & 3.92 & 1.40 \\
10.0 & 5\,10$^{3}$ & -9.5 & -9.3 & -3.5 & 1.74 & 2.12 & 1.22 \\
10.0 & 1\,10$^{4}$ & -9.7 & -9.3 & -3.5 & 1.56 & 1.44 & 0.92 \\
10.0 & 5\,10$^{4}$ & -10.1 & -9.2 & -3.4 & 1.79 & 0.64 & 0.36 \\
10.0 & 1\,10$^{5}$ & -10.2 & -9.1 & -3.4 & 1.99 & 0.48 & 0.24 \\
10.0 & 5\,10$^{5}$ & -10.5 & -9.0 & -3.4 & 2.17 & 0.23 & 0.10 \\
10.0 & 1\,10$^{6}$ & -10.7 & -9.0 & -3.4 & 2.19 & 0.16 & 0.07 \\
16.0 & 1\,10$^{3}$ & -8.6 & -8.4 & -3.5 & 10.58 & 11.99 & 1.13 \\
16.0 & 5\,10$^{3}$ & -9.1 & -8.6 & -3.4 & 6.79 & 5.78 & 0.85 \\
16.0 & 1\,10$^{4}$ & -9.3 & -8.5 & -3.4 & 6.63 & 4.18 & 0.63 \\
16.0 & 5\,10$^{4}$ & -9.7 & -8.7 & -3.4 & 5.79 & 2.05 & 0.35 \\
16.0 & 1\,10$^{5}$ & -9.9 & -8.8 & -3.4 & 5.01 & 1.55 & 0.31 \\
16.0 & 5\,10$^{5}$ & -10.2 & -9.0 & -3.3 & 3.97 & 0.72 & 0.18 \\
16.0 & 1\,10$^{6}$ & -10.4 & -8.7 & -3.3 & 4.03 & 0.49 & 0.12 \\
26.0 & 1\,10$^{3}$ & -8.4 & -8.2 & -3.4 & 12.46 & 14.49 & 1.16 \\
26.0 & 5\,10$^{3}$ & -8.9 & -8.3 & -3.4 & 9.03 & 8.66 & 0.96 \\
26.0 & 1\,10$^{4}$ & -9.2 & -8.3 & -3.4 & 9.03 & 7.03 & 0.78 \\
26.0 & 5\,10$^{4}$ & -9.6 & -8.5 & -3.4 & 8.94 & 4.13 & 0.46 \\
26.0 & 1\,10$^{5}$ & -9.7 & -8.6 & -3.4 & 8.26 & 3.21 & 0.39 \\
26.0 & 5\,10$^{5}$ & -10.1 & -8.9 & -3.3 & 6.74 & 1.53 & 0.23 \\
26.0 & 1\,10$^{6}$ & -10.2 & -8.7 & -3.3 & 6.86 & 1.05 & 0.15 \\
40.0 & 1\,10$^{3}$ & -8.3 & -8.1 & -3.4 & 13.15 & 16.22 & 1.23 \\
40.0 & 5\,10$^{3}$ & -8.9 & -8.2 & -3.4 & 10.14 & 10.78 & 1.06 \\
40.0 & 1\,10$^{4}$ & -9.1 & -8.2 & -3.4 & 10.38 & 9.44 & 0.91 \\
40.0 & 5\,10$^{4}$ & -9.5 & -8.4 & -3.3 & 11.17 & 6.46 & 0.58 \\
40.0 & 1\,10$^{5}$ & -9.7 & -8.6 & -3.3 & 10.92 & 5.20 & 0.48 \\
40.0 & 5\,10$^{5}$ & -10.0 & -8.9 & -3.3 & 9.75 & 2.58 & 0.27 \\
40.0 & 1\,10$^{6}$ & -10.2 & -8.7 & -3.3 & 10.02 & 1.78 & 0.18 \\
50.0 & 1\,10$^{3}$ & -8.3 & -8.1 & -3.4 & 13.41 & 17.13 & 1.28 \\
50.0 & 5\,10$^{3}$ & -8.8 & -8.2 & -3.4 & 10.62 & 11.84 & 1.11 \\
50.0 & 1\,10$^{4}$ & -9.0 & -8.2 & -3.4 & 11.01 & 10.68 & 0.97 \\
50.0 & 5\,10$^{4}$ & -9.5 & -8.4 & -3.3 & 12.26 & 7.85 & 0.64 \\
50.0 & 1\,10$^{5}$ & -9.6 & -8.6 & -3.3 & 12.26 & 6.44 & 0.53 \\
50.0 & 5\,10$^{5}$ & -10.0 & -8.8 & -3.3 & 11.49 & 3.28 & 0.29 \\
50.0 & 1\,10$^{6}$ & -10.1 & -8.7 & -3.3 & 11.88 & 2.28 & 0.19 \\
\hline
\multicolumn{8}{c}{ \Gn = 50}\\
\hline
2.0 & 1\,10$^{3}$ & -11.4 & -11.9 & -8.0 & 2.55 & 17.66 & 6.93 \\
2.0 & 5\,10$^{3}$ & -11.7 & -11.0 & -6.6 & 1.14 & 0.47 & 0.41 \\
2.0 & 1\,10$^{4}$ & -11.7 & -10.7 & -6.0 & 0.68 & 0.15 & 0.23 \\
2.0 & 5\,10$^{4}$ & -11.5 & -10.1 & -4.7 & 0.18 & 0.03 & 0.14 \\
2.0 & 1\,10$^{5}$ & -11.4 & -9.9 & -4.3 & 0.12 & 0.01 & 0.12 \\
2.0 & 5\,10$^{5}$ & -10.8 & -9.2 & -3.7 & 0.23 & 0.02 & 0.07 \\
2.0 & 1\,10$^{6}$ & -9.4 & -7.1 & -3.6 & 0.85 & 0.02 & 0.02 \\
4.0 & 1\,10$^{3}$ & -11.5 & -11.0 & -5.7 & 0.14 & 0.08 & 0.58 \\
4.0 & 5\,10$^{3}$ & -11.5 & -10.3 & -4.3 & 0.08 & 0.01 & 0.14 \\
4.0 & 1\,10$^{4}$ & -11.4 & -10.1 & -4.0 & 0.12 & 0.01 & 0.11 \\
4.0 & 5\,10$^{4}$ & -11.2 & -9.5 & -3.7 & 0.32 & 0.02 & 0.06 \\
4.0 & 1\,10$^{5}$ & -11.2 & -9.4 & -3.6 & 0.43 & 0.02 & 0.04 \\
4.0 & 5\,10$^{5}$ & -11.0 & -9.0 & -3.5 & 0.73 & 0.02 & 0.02 \\
4.0 & 1\,10$^{6}$ & -10.2 & -7.5 & -3.5 & 1.16 & 0.01 & 0.01 \\
6.0 & 1\,10$^{3}$ & -11.2 & -10.5 & -4.9 & 0.17 & 0.07 & 0.45 \\
6.0 & 5\,10$^{3}$ & -10.9 & -10.0 & -3.8 & 0.23 & 0.06 & 0.27 \\
6.0 & 1\,10$^{4}$ & -10.9 & -9.8 & -3.7 & 0.33 & 0.06 & 0.18 \\
6.0 & 5\,10$^{4}$ & -11.0 & -9.4 & -3.5 & 0.62 & 0.05 & 0.08 \\
6.0 & 1\,10$^{5}$ & -11.0 & -9.3 & -3.5 & 0.76 & 0.04 & 0.06 \\
6.0 & 5\,10$^{5}$ & -11.0 & -9.0 & -3.4 & 1.02 & 0.03 & 0.03 \\
6.0 & 1\,10$^{6}$ & -10.2 & -7.9 & -3.4 & 1.25 & 0.02 & 0.02 \\
10.0 & 1\,10$^{3}$ & -10.1 & -9.7 & -4.0 & 1.11 & 0.85 & 0.77 \\
10.0 & 5\,10$^{3}$ & -9.9 & -9.5 & -3.6 & 1.02 & 0.95 & 0.93 \\
10.0 & 1\,10$^{4}$ & -9.9 & -9.5 & -3.5 & 0.97 & 0.73 & 0.75 \\
10.0 & 5\,10$^{4}$ & -10.3 & -9.3 & -3.4 & 1.05 & 0.36 & 0.34 \\
10.0 & 1\,10$^{5}$ & -10.4 & -9.3 & -3.4 & 1.12 & 0.27 & 0.24 \\
10.0 & 5\,10$^{5}$ & -10.7 & -9.2 & -3.4 & 1.27 & 0.12 & 0.10 \\
10.0 & 1\,10$^{6}$ & -10.5 & -8.0 & -3.4 & 1.55 & 0.08 & 0.05 \\
16.0 & 1\,10$^{3}$ & -8.8 & -8.6 & -3.6 & 9.36 & 10.90 & 1.16 \\
16.0 & 5\,10$^{3}$ & -9.2 & -8.8 & -3.5 & 4.95 & 4.54 & 0.92 \\
16.0 & 1\,10$^{4}$ & -9.4 & -8.7 & -3.4 & 4.67 & 3.21 & 0.69 \\
16.0 & 5\,10$^{4}$ & -9.8 & -8.9 & -3.4 & 3.76 & 1.58 & 0.42 \\
16.0 & 1\,10$^{5}$ & -9.9 & -9.0 & -3.4 & 3.20 & 1.19 & 0.37 \\
16.0 & 5\,10$^{5}$ & -10.2 & -9.0 & -3.4 & 2.60 & 0.53 & 0.20 \\
16.0 & 1\,10$^{6}$ & -10.2 & -8.0 & -3.4 & 2.98 & 0.34 & 0.11 \\
26.0 & 1\,10$^{3}$ & -8.5 & -8.3 & -3.5 & 12.16 & 14.10 & 1.16 \\
26.0 & 5\,10$^{3}$ & -9.0 & -8.4 & -3.4 & 7.76 & 7.55 & 0.97 \\
26.0 & 1\,10$^{4}$ & -9.2 & -8.4 & -3.4 & 7.55 & 6.07 & 0.80 \\
26.0 & 5\,10$^{4}$ & -9.6 & -8.6 & -3.4 & 6.94 & 3.61 & 0.52 \\
26.0 & 1\,10$^{5}$ & -9.7 & -8.7 & -3.4 & 6.15 & 2.81 & 0.46 \\
26.0 & 5\,10$^{5}$ & -10.1 & -8.9 & -3.4 & 4.94 & 1.26 & 0.26 \\
26.0 & 1\,10$^{6}$ & -10.2 & -8.3 & -3.3 & 5.29 & 0.81 & 0.15 \\
40.0 & 1\,10$^{3}$ & -8.3 & -8.2 & -3.4 & 12.97 & 15.95 & 1.23 \\
40.0 & 5\,10$^{3}$ & -8.9 & -8.3 & -3.4 & 8.88 & 9.64 & 1.08 \\
40.0 & 1\,10$^{4}$ & -9.1 & -8.3 & -3.4 & 8.88 & 8.42 & 0.95 \\
40.0 & 5\,10$^{4}$ & -9.5 & -8.5 & -3.4 & 9.11 & 5.91 & 0.65 \\
40.0 & 1\,10$^{5}$ & -9.6 & -8.7 & -3.4 & 8.64 & 4.76 & 0.55 \\
40.0 & 5\,10$^{5}$ & -10.0 & -8.9 & -3.3 & 7.58 & 2.20 & 0.29 \\
40.0 & 1\,10$^{6}$ & -10.1 & -8.3 & -3.3 & 8.18 & 1.43 & 0.17 \\
50.0 & 1\,10$^{3}$ & -8.3 & -8.1 & -3.4 & 13.25 & 16.90 & 1.28 \\
50.0 & 5\,10$^{3}$ & -8.8 & -8.3 & -3.4 & 9.35 & 10.66 & 1.14 \\
50.0 & 1\,10$^{4}$ & -9.0 & -8.3 & -3.4 & 9.47 & 9.61 & 1.01 \\
50.0 & 5\,10$^{4}$ & -9.5 & -8.5 & -3.4 & 10.14 & 7.29 & 0.72 \\
50.0 & 1\,10$^{5}$ & -9.6 & -8.6 & -3.3 & 9.92 & 5.99 & 0.60 \\
50.0 & 5\,10$^{5}$ & -10.0 & -8.9 & -3.3 & 9.16 & 2.84 & 0.31 \\
50.0 & 1\,10$^{6}$ & -10.1 & -8.5 & -3.3 & 9.77 & 1.85 & 0.19 \\
\hline
\multicolumn{8}{c}{ \Gn = 100}\\
\hline
2.0 & 1\,10$^{3}$ & -11.5 & -12.6 & -8.6 & 2.27 & 69.47 & 30.54 \\
2.0 & 5\,10$^{3}$ & -11.7 & -11.5 & -7.4 & 1.97 & 2.80 & 1.42 \\
2.0 & 1\,10$^{4}$ & -11.8 & -11.1 & -6.8 & 1.46 & 0.85 & 0.59 \\
2.0 & 5\,10$^{4}$ & -11.7 & -10.5 & -5.5 & 0.47 & 0.11 & 0.23 \\
2.0 & 1\,10$^{5}$ & -11.6 & -10.3 & -5.0 & 0.23 & 0.04 & 0.19 \\
2.0 & 5\,10$^{5}$ & -10.5 & -9.5 & -4.0 & 0.12 & 0.05 & 0.38 \\
2.0 & 1\,10$^{6}$ & -10.7 & -9.2 & -3.8 & 0.15 & 0.01 & 0.07 \\
4.0 & 1\,10$^{3}$ & -11.6 & -11.3 & -6.3 & 0.24 & 0.27 & 1.13 \\
4.0 & 5\,10$^{3}$ & -11.7 & -10.6 & -4.7 & 0.07 & 0.01 & 0.18 \\
4.0 & 1\,10$^{4}$ & -11.6 & -10.3 & -4.3 & 0.08 & 0.01 & 0.12 \\
4.0 & 5\,10$^{4}$ & -11.4 & -9.7 & -3.8 & 0.21 & 0.01 & 0.07 \\
4.0 & 1\,10$^{5}$ & -11.3 & -9.5 & -3.7 & 0.30 & 0.01 & 0.05 \\
4.0 & 5\,10$^{5}$ & -10.7 & -9.1 & -3.6 & 0.53 & 0.03 & 0.06 \\
4.0 & 1\,10$^{6}$ & -11.0 & -9.0 & -3.5 & 0.60 & 0.01 & 0.02 \\
6.0 & 1\,10$^{3}$ & -11.3 & -10.7 & -5.1 & 0.13 & 0.06 & 0.47 \\
6.0 & 5\,10$^{3}$ & -11.1 & -10.1 & -3.9 & 0.16 & 0.04 & 0.26 \\
6.0 & 1\,10$^{4}$ & -11.0 & -9.9 & -3.8 & 0.23 & 0.04 & 0.18 \\
6.0 & 5\,10$^{4}$ & -11.1 & -9.5 & -3.6 & 0.45 & 0.04 & 0.08 \\
6.0 & 1\,10$^{5}$ & -11.1 & -9.4 & -3.5 & 0.56 & 0.03 & 0.06 \\
6.0 & 5\,10$^{5}$ & -10.8 & -9.1 & -3.5 & 0.80 & 0.04 & 0.05 \\
6.0 & 1\,10$^{6}$ & -11.1 & -9.0 & -3.4 & 0.83 & 0.02 & 0.02 \\
10.0 & 1\,10$^{3}$ & -10.3 & -9.8 & -4.2 & 0.90 & 0.61 & 0.68 \\
10.0 & 5\,10$^{3}$ & -10.0 & -9.6 & -3.6 & 0.83 & 0.64 & 0.76 \\
10.0 & 1\,10$^{4}$ & -10.1 & -9.5 & -3.6 & 0.82 & 0.52 & 0.64 \\
10.0 & 5\,10$^{4}$ & -10.3 & -9.4 & -3.5 & 0.88 & 0.28 & 0.32 \\
10.0 & 1\,10$^{5}$ & -10.5 & -9.4 & -3.4 & 0.93 & 0.21 & 0.23 \\
10.0 & 5\,10$^{5}$ & -10.7 & -9.2 & -3.4 & 1.01 & 0.11 & 0.11 \\
10.0 & 1\,10$^{6}$ & -10.8 & -9.2 & -3.4 & 0.99 & 0.06 & 0.06 \\
16.0 & 1\,10$^{3}$ & -8.8 & -8.7 & -3.7 & 8.80 & 10.47 & 1.19 \\
16.0 & 5\,10$^{3}$ & -9.2 & -8.9 & -3.5 & 4.15 & 4.05 & 0.98 \\
16.0 & 1\,10$^{4}$ & -9.4 & -8.8 & -3.5 & 3.85 & 2.85 & 0.74 \\
16.0 & 5\,10$^{4}$ & -9.8 & -9.0 & -3.4 & 3.03 & 1.40 & 0.46 \\
16.0 & 1\,10$^{5}$ & -10.0 & -9.0 & -3.4 & 2.59 & 1.06 & 0.41 \\
16.0 & 5\,10$^{5}$ & -10.3 & -9.1 & -3.4 & 2.15 & 0.47 & 0.22 \\
16.0 & 1\,10$^{6}$ & -10.4 & -9.1 & -3.4 & 2.13 & 0.29 & 0.14 \\
26.0 & 1\,10$^{3}$ & -8.5 & -8.3 & -3.5 & 12.27 & 14.21 & 1.16 \\
26.0 & 5\,10$^{3}$ & -9.0 & -8.4 & -3.4 & 7.31 & 7.17 & 0.98 \\
26.0 & 1\,10$^{4}$ & -9.2 & -8.4 & -3.4 & 7.04 & 5.74 & 0.82 \\
26.0 & 5\,10$^{4}$ & -9.6 & -8.6 & -3.4 & 6.23 & 3.45 & 0.55 \\
26.0 & 1\,10$^{5}$ & -9.7 & -8.8 & -3.4 & 5.44 & 2.68 & 0.49 \\
26.0 & 5\,10$^{5}$ & -10.0 & -9.0 & -3.4 & 4.41 & 1.18 & 0.27 \\
26.0 & 1\,10$^{6}$ & -10.2 & -8.9 & -3.4 & 4.47 & 0.73 & 0.16 \\
40.0 & 1\,10$^{3}$ & -8.3 & -8.2 & -3.4 & 13.13 & 16.13 & 1.23 \\
40.0 & 5\,10$^{3}$ & -8.9 & -8.3 & -3.4 & 8.45 & 9.27 & 1.10 \\
40.0 & 1\,10$^{4}$ & -9.1 & -8.3 & -3.4 & 8.37 & 8.09 & 0.97 \\
40.0 & 5\,10$^{4}$ & -9.5 & -8.5 & -3.4 & 8.38 & 5.76 & 0.69 \\
40.0 & 1\,10$^{5}$ & -9.6 & -8.7 & -3.4 & 7.86 & 4.64 & 0.59 \\
40.0 & 5\,10$^{5}$ & -10.0 & -8.9 & -3.3 & 6.96 & 2.10 & 0.30 \\
40.0 & 1\,10$^{6}$ & -10.1 & -8.9 & -3.3 & 7.22 & 1.32 & 0.18 \\
50.0 & 1\,10$^{3}$ & -8.3 & -8.2 & -3.4 & 13.42 & 17.11 & 1.27 \\
50.0 & 5\,10$^{3}$ & -8.8 & -8.3 & -3.4 & 8.91 & 10.28 & 1.15 \\
50.0 & 1\,10$^{4}$ & -9.0 & -8.3 & -3.4 & 8.94 & 9.28 & 1.04 \\
50.0 & 5\,10$^{4}$ & -9.4 & -8.5 & -3.4 & 9.40 & 7.14 & 0.76 \\
50.0 & 1\,10$^{5}$ & -9.6 & -8.7 & -3.4 & 9.10 & 5.88 & 0.65 \\
50.0 & 5\,10$^{5}$ & -9.9 & -8.9 & -3.3 & 8.51 & 2.71 & 0.32 \\
50.0 & 1\,10$^{6}$ & -10.1 & -8.8 & -3.3 & 8.92 & 1.72 & 0.19 \\
\end{longtable}
\tablefoot{\tablefoottext{a}{logarithmic relative abundance
    X(mol)$=\log(N(\rm mol)/N({\rm H}_2))$}}
}
\end{appendix}

\end{document}